\documentclass[%
 reprint,
 superscriptaddress,
 amsmath,amssymb,
 aps,
 prl,
]{revtex4-2}

\usepackage{physics}
\usepackage{graphicx}
\usepackage{dcolumn}
\usepackage{bm}
\usepackage{booktabs} 

\usepackage{xcolor}

\begin{document}

\title{Broken neural scaling laws in materials science}

\author{Max Großmann}
\email{max.grossmann@tu-ilmenau.de}
\affiliation{Institute of Physics and Institute of Micro- and Nanotechnologies, Technische Universit\"at Ilmenau, 98693 Ilmenau, Germany}

\author{Malte Grunert}
\affiliation{Institute of Physics and Institute of Micro- and Nanotechnologies, Technische Universit\"at Ilmenau, 98693 Ilmenau, Germany}

\author{Erich Runge}
\affiliation{Institute of Physics and Institute of Micro- and Nanotechnologies, Technische Universit\"at Ilmenau, 98693 Ilmenau, Germany}

\date{\today}

\begin{abstract}
    In materials science, data are scarce and expensive to generate, whether computationally or experimentally.
    Therefore, it is crucial to identify how model performance scales with dataset size and model capacity to distinguish between data- and model-limited regimes.
    Neural scaling laws provide a framework for quantifying this behavior and guide the design of materials datasets and machine learning architectures.
    Here, we investigate neural scaling laws for a paradigmatic materials science task: predicting the dielectric function of metals, a high-dimensional response that governs how solids interact with light.
    Using over 200,000 dielectric functions from high-throughput ab initio calculations, we study two multi-objective graph neural networks trained to predict the frequency-dependent complex interband dielectric function and the Drude frequency.
    We observe broken neural scaling laws with respect to dataset size, whereas scaling with the number of model parameters follows a simple power law that rapidly saturates.
\end{abstract}

\maketitle

Neural scaling laws (NSLs) quantify how the performance of machine learning (ML) models scales with respect to training data, model parameters, and architectural choices, and are widely used to guide model development across domains, including computer vision and large language models (LLMs) \cite{Hestness2017, Rosenfeld2019, Kaplan2020, Zhai2021, Hoffmann2022, Grattafiori2024}.
While ML is revolutionizing materials science, research into NSLs for materials science graph neural networks (GNNs) is just beginning in materials discovery \cite{Merchant2023}, deep chemical models \cite{Frey2023}, and machine-learned interatomic potentials (MLIPs) \cite{Ngo2025}.
This is unfortunate, as NSLs quantify how predictive performance scales with data and model capacity, thereby guiding the efficient use of computational and data resources.

NSLs are generally reported to follow simple power-law behavior \cite{Hestness2017, Kaplan2020, Hoffmann2022, Frey2023, Grattafiori2024, Ngo2025}.
However, so-called broken neural scaling laws (BNSLs)---characterized by deviations from simple power laws, such as slope changes, saturation, or unexpected steepening---have been reported for a number of tasks and learning settings by Caballero et al.~\cite{Caballero2022}, but not yet within the context of materials science.
Here, we report on the emergence of BNSLs in data scaling in the field of materials science.
This finding is significant because data in materials science are both scarce and expensive to generate, whether computationally or experimentally.
Consequently, identifying where data scaling saturates or steepens is crucial for distinguishing between data- and model-limited regimes, and for guiding the design of future materials datasets and ML architectures.

In this study, we investigate NSLs for a generic and paradigmatic materials science task: predicting the dielectric function $\varepsilon(\omega)$ of metals, i.e., a high-dimensional physical response that directly encodes how solids interact with light and underpins a wide range of optical technologies.
From an ML perspective, learning $\varepsilon(\omega)$ of metals is challenging, as it constitutes a multi-objective task, requiring the prediction of both interband transitions and the intraband response characterized by a Drude frequency \cite{Dressel2002}.
Most importantly, this task highlights the fundamental constraints that govern data scaling in materials science.
While recent GNNs made the efficient prediction of optical spectra for semiconductors and insulators possible, their training datasets remain limited in size and accuracy \cite{Ibrahim2024, Grunert2024, Hung2024}, largely due to the computational cost of climbing the Jacob's Ladder of optoelectronic properties \cite{Grunert2025}.
Metals, by contrast, provide a physically well-controlled regime in which it is feasible to generate accurate optical data on a large scale using comparatively lower-level ab initio methods.
This is facilitated by strong electronic screening in metals, which suppresses excitonic effects and often allows for quantitative agreement with experiments within the independent-particle approximation (IPA) augmented by a Drude term \cite{Marini2001, Prandini2019Photo, Barker2022}.
Consequently, learning the optical properties of metals provides an attractive combination of technological relevance and computational tractability in terms of accuracy and scalability, making it well suited for probing NSLs in the data-limited regime of materials science.

Based on these considerations and to obtain meaningful scaling laws with respect to dataset size $D$, we generated a dataset containing 201,361 dielectric functions of metals using high-throughput ab initio calculations. 
This dataset exceeds existing datasets \cite{Grunert2024, Hung2024, Hsu2025, Grunert2025, Grunert2026} for dielectric functions of semiconductors and insulators by approximately an order of magnitude.
To probe scaling behavior with respect to model parameter count $N$, we employ architectures based on our previous work \cite{Grunert2024, Grunert2025, Grunert2026}, which were deliberately streamlined to minimize architectural degrees of freedom and enable controlled variation of $N$.
Specifically, we use two rotationally invariant multi-objective GNNs based on two- and three-body interactions, trained to simultaneously predict the frequency-dependent complex interband dielectric function and the Drude frequency.
Comparing the two- and three-body variants after detailed architecture optimization allows us to directly assess how increasing body order affects both predictive performance and NSLs.

\section*{Results}

In the following, we analyze the scaling behavior of GNNs that predict $\varepsilon(\omega)$ of metals from structural information alone.
As detailed in the Methods section, the dielectric response can be separated into a complex interband contribution, $\varepsilon_\mathrm{inter}(\omega)$, and a Drude term proportional to the square of the Drude frequency, $\omega_\mathrm{D}$.
Both are in general tensors \cite{Hsu2025}, but we restrict ourselves to rotationally invariant averages \cite{Carrico2024, Grunert2024, Grunert2025} (Greek indices refer to the Cartesian components): 
$\overline{\varepsilon}_\mathrm{inter}(\omega) = \mathrm{Tr}[\varepsilon_{\mathrm{inter},\alpha\beta}(\omega)]/3$
and $\overline{\omega}_\mathrm{D}^{\,2} = \mathrm{Tr}[\omega_{\mathrm{D},\alpha\beta}^2]/3$ (consistent with the quadratic appearance of $\omega_\mathrm{D}$ in the intraband contributions), respectively.

\subsection{Dataset}

Extracting reliable NSLs requires large and robust datasets, which are scarce in materials science, particularly for optical properties.
Even the largest existing datasets contain fewer than 30,000 spectra \cite{Grunert2026}, which is insufficient to systematically probe scaling behavior across multiple orders of magnitude in dataset size.

To address this limitation, we generated a large-scale dataset of dielectric functions for 205,224 intermetallic compounds using high-throughput ab initio calculations based on structures filtered from the \textsc{Alexandria} database \cite{Schmidt2023, Schmidt2024} (downloaded March 2024).
All calculations were automated using custom in-house workflows to ensure full reproducibility.
Converged calculations for $\overline{\varepsilon}_\mathrm{inter}(\omega)$ and $\overline{\omega}_\mathrm{D}$ were obtained for 201,361 materials, while for the remaining 3,863 materials the IPA part of the workflow could not be converged within the available computational budget.
The structural filtering criteria, calculation methodology, and design of the high-throughput workflow are described in the Methods section.
Supplementary Note 1 provides additional statistics on the prevalence of elements and the distribution of unit cell sizes across the dataset.
The generated dataset and workflow code are provided in the Data and Code Availability statements, respectively.

To validate the quality of the employed calculation methodology and, consequently, of the resulting dataset, we compared the dielectric functions of $27$ elemental metals computed with our workflow against experimental data (see Supplementary Note 2).
Overall, we find good agreement between the calculated and experimental spectra in both peak positions and spectral shape, consistent with the results of Prandini et al.~\cite{Prandini2019Photo}.

\subsection{Model setup}

To investigate NSLs in a controlled setting, we use invariant GNNs with deliberately constrained architectural degrees of freedom.
Building on our previous \textsc{OptiMate} models for optical spectra of semiconductors and insulators \cite{Grunert2024, Grunert2025, Grunert2026}, we adapt the architecture to the present multi-objective task and introduce two variants, \textsc{OptiMetal2B} and \textsc{OptiMetal3B}.
\textsc{OptiMetal2B} follows the original two-body message-passing formulation of \textsc{OptiMate}, whereas \textsc{OptiMetal3B} extends this framework by explicitly incorporating three-body interactions, building on our recent \textsc{OptiMate3B} developments \cite{Grunert2026}.
Both models share the same design philosophy and training protocol, differing only in their effective body order (two-body versus three-body interactions) and, for selected variants, their message-passing formulation (\textsc{Crystal Graph Convolution} \cite{Xie2018}, CGC, versus \textsc{Transformer Convolution} \cite{Thekumparampil2018}, TC).
The basis for both \textsc{OptiMetal2B} and \textsc{OptiMetal3B} is a set of invariant input features (atom types, bond lengths, and bond angles), which enable a controlled assessment of the effect of body order on model performance and scaling behavior.
Extending the following NSL analysis to equivariant GNNs is a natural direction for future work.
A detailed description of the model architectures is provided in Supplementary Note 3.

Prior to the NSL analysis, the architectural components and hyperparameters were systematically optimized at a fixed model width (Supplementary Note 4), ensuring that the resulting scaling laws reflect near-optimal model performance rather than architectural inefficiencies.
The resulting models accurately predict $\overline{\varepsilon}_\mathrm{inter}(\omega)$, $\overline{\omega}_\mathrm{D}$, and material colors on the held-out test set.
Detailed performance metrics and representative predictions are reported in Supplementary Note 5.

\subsection{Neural scaling laws}

\begin{figure}
    \centering
    \includegraphics{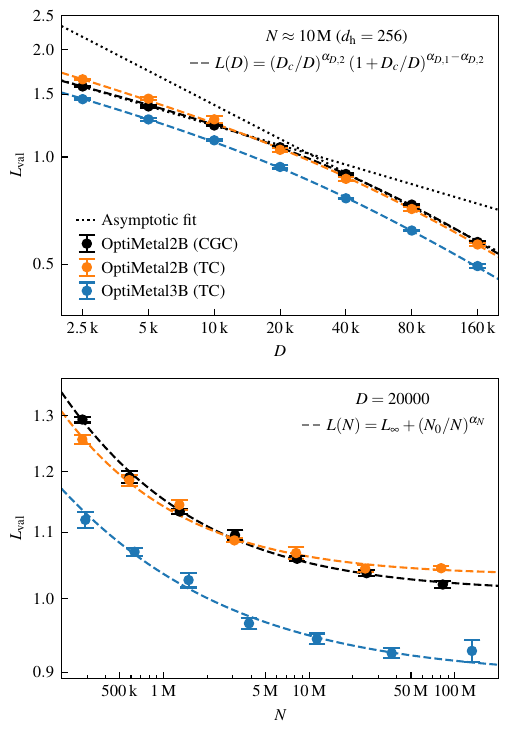}
    \caption{
    \textbf{1D NSLs for dataset and parameter scaling for \textsc{OptiMetal2B} and \textsc{OptiMetal3B}.}
    The upper panel shows the validation loss $L_\mathrm{val}$ as a function of dataset size $D$ at fixed parameter count $N$ for \textsc{OptiMetal2B} and \textsc{OptiMetal3B}, and vice versa in the lower panel.
    Dots represent $L_\mathrm{val}$ averaged over three random model initializations, and error bars indicate the standard deviation.
    The dashed lines show the corresponding best NSL fits, whose functional form is given in the panel legends (see Methods for details).
    We investigate the scaling behavior of \textsc{OptiMetal2B} using CGC \cite{Xie2018} and TC \cite{Thekumparampil2018} message passing, as both yield nearly identical performance after the architecture optimization (see Supplementary Note 4).
    Fit parameters, including scaling exponents, are given in Tab.~\ref{tab:scaling_laws}.
    In the top panel, we show two conventional NSLs (simple power laws) fitted to the first and last three data points of \textsc{OptiMetal2B} (CGC) (dotted black lines, annotated as "Asymptotic fit"), to highlight the "brokenness" of the neural scaling behavior.
    }
    \label{fig:1dscaling}
\end{figure}

Having established accurate baseline \textsc{OptiMetal2B} and \textsc{OptiMetal3B} models for the multi-objective task at hand, we next quantify how performance scales with dataset size $D$ and parameter count $N$.
To assess the sensitivity of NSLs to the message-passing formulation, we compare \textsc{OptiMetal2B} variants with CGC \cite{Xie2018} and TC \cite{Thekumparampil2018} message passing, as both yield nearly identical performance after the architecture optimization (see Supplementary Note 4).
This allows us to isolate the effect of the message-passing formulation on the scaling behavior.

For data scaling, we randomly subsampled the training set into seven subsets containing 2,500, 5,000, 10,000, 20,000, 40,000, 80,000, and 160,000 materials, respectively, and used these subsets consistently throughout the subsequent NSL analysis.
To ensure that these subsets remained representative, we verified the elemental prevalence and unit cell size distributions for each subset and found no significant statistical deviations (see Supplementary Note 6).
The parameter count $N$ was controlled by varying the hidden dimension $d_\mathrm{h}$ (cf.~Supplementary Note 3) in powers of two from 16 to 1,024, corresponding to models with approximately $N\approx 100\,\mathrm{k}$--$ 100\,\mathrm{M}$ parameters (only trainable parameters were included in $N$).

First, we analyze one-dimensional (1D) NSLs, i.e., we investigate $L(D)$ while holding $N$ constant and vice versa.
The loss function $L$ is defined as the sum of mean absolute errors (MAE) for the interband dielectric function and the Drude frequency (see Methods).
In line with the architecture optimization described in Supplementary Note 4, we set $d_\mathrm{h}=256$ ($N\approx 10\,\mathrm{M}$) for data scaling $L(D)$, and fixed $D=20{,}000$ for parameter scaling $L(N)$. 

Before presenting the 1D NSLs, we describe how the functional forms used to fit the validation loss, $L_\mathrm{val}$, were selected, as this choice is central to identifying BNSLs.
Inspired by previous studies on NSLs \cite{Hestness2017, Kaplan2020, Hoffmann2022, Caballero2022, Frey2023, Ngo2025}, we evaluated four candidate functions for each model and 1D NSL, see Eqs.~(\ref{eq:pl})--\ref{eq:broken_amp}) below: a simple power law, a power law with a saturation floor, and two smoothly broken power laws, one without and one with an adjustable amplitude.
To determine the function that best describes the scaling behavior, we compared all fitted functional forms using the Akaike information criterion corrected for small sample sizes (AICc) \cite{Burnham2002}.
The AICc balances fit quality with the number of parameters used to achieve it by penalizing overly complex functions that would otherwise overfit the data.
The fit with the lowest AICc value is considered the preferred functional form.
The functional forms of the 1D NSLs are defined in the Methods section, and Supplementary Note 7 reports the corresponding AICc values for all tested fits.

\begin{table}
    \centering
    \caption{
    \textbf{Fit parameters of the 1D NSLs obtained from the AICc-selected functional forms described in the Methods section.}
    The reported parameters are obtained from nonlinear least squares fits to validation losses averaged over three random model initializations.
    For data scaling, the validation loss is best described by the smoothly broken power law given in Eq.~(\ref{eq:broken}), yielding the low- and high-data exponents $\alpha_{D,1}$ and $\alpha_{D,2}$ as well as the crossover scale $D_c$.
    For parameter scaling, the preferred NSL is the power law with a saturation floor in Eq.~(\ref{eq:pl_floor}), characterized by the exponent $\alpha_N$, the characteristic parameter count $N_0$, and the irreducible loss $L_\infty$.
    }
    \vspace*{2mm}
    \begin{tabular}{l@{\hspace{1em}}c@{\hspace{1em}}c@{\hspace{1em}}c@{\hspace{1em}}}
    \toprule
    ~~~Data scaling & $\alpha_{D,1}$ & $\alpha_{D,2}$ & $D_c$ \\
    \midrule
    \textsc{OptiMetal2B} (CGC) & $0.15$ & $0.42$ & $10^{\,4.72}$ \\
    \textsc{OptiMetal2B} (TC) & $0.18$ & $0.39$ & $10^{\,4.62}$ \\
    \textsc{OptiMetal3B} (TC) & $0.17$ & $0.38$ & $10^{\,4.43}$ \\
    \bottomrule
    \toprule
    ~~~Parameter scaling & $\alpha_{N}$ & $N_0$ & $L_\infty$ \\
    \midrule
    \textsc{OptiMetal2B} (CGC) & $0.53$ & $10^{\,4.40}$ & $1.01$ \\
    \textsc{OptiMetal2B} (TC) & $0.58$ & $10^{\,4.33}$ & $1.03$ \\
    \textsc{OptiMetal3B} (TC) & $0.41$ & $10^{\,3.95}$ & $0.89$ \\
    \bottomrule
    \end{tabular}
    \label{tab:scaling_laws}
\end{table}

Based on this analysis, we found that 1D data scaling is best described by a smoothly broken power law without an adjustable amplitude, whereas 1D parameter scaling is best captured by a power law with a saturation floor.
Figure~\ref{fig:1dscaling} summarizes the resulting 1D NSLs for data scaling $L(D)$ (top) and parameter scaling $L(N)$ (bottom), with the corresponding functional forms indicated in the figure legends and detailed in the Methods section.
Across all architectures, similar data and parameter NSLs are found, with fit parameters listed in Tab.~\ref{tab:scaling_laws}.

For data scaling, \textsc{OptiMetal2B} (CGC and TC) and \textsc{OptiMetal3B} exhibit similar scaling exponents in the low-data regime, with $\alpha_{D,1}\approx0.15$--$0.18$, suggesting that all models extract information from small datasets with comparable efficiency.
Above the breakpoint $D_c$ between $D=10^{\,4.43}$--$10^{\,4.72}$, the scaling slope steepens significantly to $\alpha_{D,2}=0.38$--$0.42$, indicating a shift to a regime in which additional data yields disproportionately larger performance gains.
In this high-data regime, the scaling behavior once again appears to be largely model-independent, with similar scaling exponents across all architectures.
Consistent with this, the data scaling of the TC- and CGC-based variants of \textsc{OptiMetal2B} is nearly identical.
The CGC-based variant performs slightly better in the low-data regime, although it has a slightly lower scaling exponent $\alpha_{D,1}$.
The addition of three-body interactions in \textsc{OptiMetal3B} improves performance through a constant shift of the scaling curve toward lower validation losses.
However, within the 1D NSL analysis, three-body interactions have no significant impact on the scaling exponents, a conclusion that is partly revised in the subsequent 2D NSL analysis.

The parameter scaling exhibits a qualitatively different behavior.
All models follow a simple power-law trend with exponents $\alpha_N=0.41$--$0.58$ up to approximately $5\,\mathrm{M}$ parameters, beyond which the validation loss rapidly saturates.
The validation loss with respect to the number of model parameters saturates more slowly for \textsc{OptiMetal3B} with $\alpha_N=0.41$ compared to the \textsc{OptiMetal2B} models with $\alpha_N=0.53$--$0.58$.
As with data scaling, the parameter scaling of the TC- and CGC-based variants of \textsc{OptiMetal2B} is quite similar.
However, the TC-based variant now performs slightly better in the low-parameter regime, though it saturates at a higher validation loss as $N$ increases.

Since the hidden dimension was set to $d_\mathrm{h}=256$ (i.e., $N\approx10\,\mathrm{M}$ parameters) for the 1D data scaling runs, the models operate in a strongly overparameterized regime across all dataset sizes.
In the low-data regime, however, the limited training data may not uniquely constrain the structure-property relationship, possibly allowing many parameter configurations to fit the data equally well. 
As a result, the models may exhibit a type of "best-guess" behavior, capturing broad, global trends correlated with optical response.
Meanwhile, finer spectral features remain inadequately constrained, which is consistent with the small data-scaling exponent observed in the low-data regime.

By contrast, the increase of the data-scaling exponent above $D_c$ suggests a transition to a data-efficient regime, in which larger datasets sample a broader range of structural and electronic environments, enabling the learning of more detailed and complex structure-property correlations.
In this regime, each additional material may provide increasingly informative constraints, resulting in the second, steeper branch of the scaling curve.

This interpretation raises the question of whether the observed BNSLs persist when both $D$ and $N$ are varied simultaneously.
If the BNSLs were primarily an artifact of performing data scaling with overparameterized models, then models whose capacity is well matched to the available data should exhibit simple power-law scaling.
To test this, we analyze two-dimensional (2D) NSL maps, i.e., $L(D,N)$, which allow us to disentangle the interplay between dataset size and model capacity, thereby determining whether the observed BNSLs are intrinsic to the learning task or specific to the fixed-model-size setting.

\begin{figure*}
    \centering
    \includegraphics{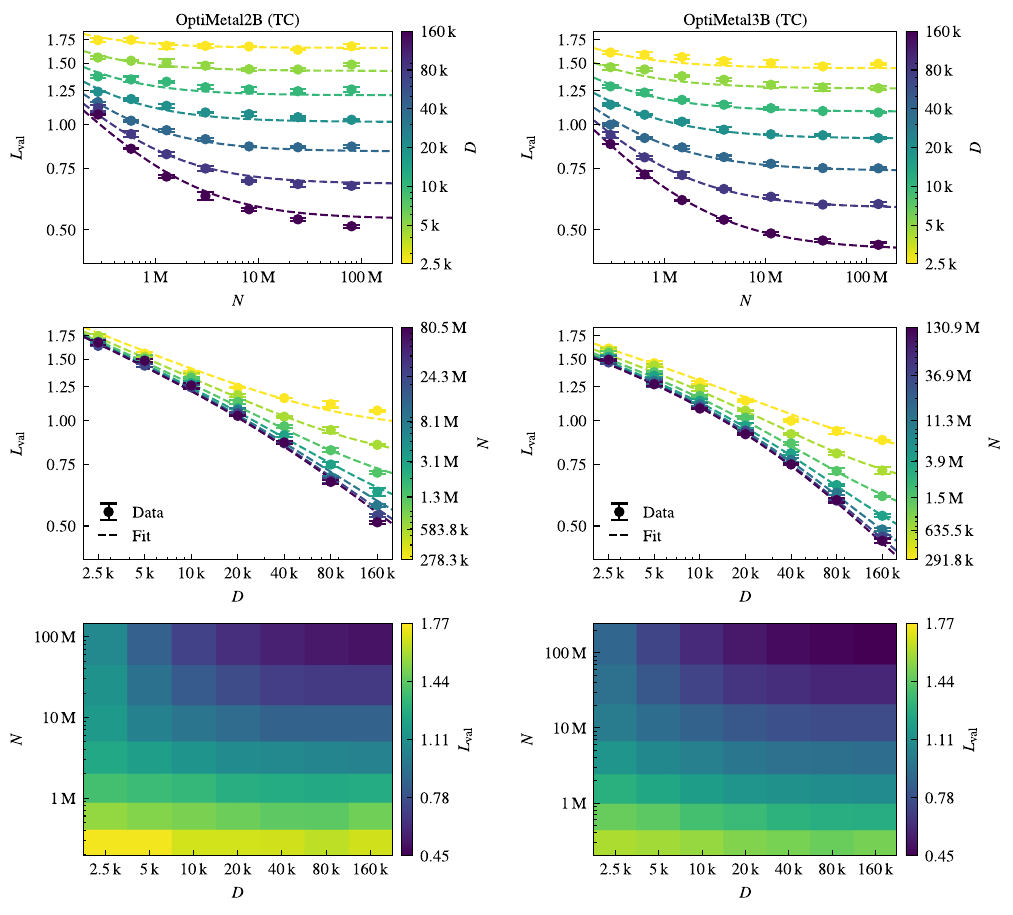}
    \caption{
    \textbf{2D NSL maps for the TC-based \textsc{OptiMetal2B} and \textsc{OptiMetal3B}.}
    The validation loss $L_\mathrm{val}$ is shown as a function of dataset size $D$ and parameter count $N$ for \textsc{OptiMetal2B} and \textsc{OptiMetal3B}.
    Dots represent $L_\mathrm{val}$ averaged over three random model initializations, and error bars indicate the standard deviation.
    The dashed lines show the corresponding best 2D NSL fits, whose functional form follows Eq.~(\ref{eq:kaplan_map}) (see Methods for details).
    Each panel provides a different illustration of the scaling behavior:
    The upper panel shows the validation loss for models with different dataset sizes $D$ as a function of parameter count $N$, as indicated by the color bar.
    The middle panel swaps the roles of dataset size $D$ and parameter count $N$.
    The lower panel provides a full 2D representation of the NSLs, though it omits the fits for clarity.
    Fit parameters, including scaling exponents, are given in Tab.~\ref{tab:scaling_maps}.
    }
    \label{fig:3dscaling}
\end{figure*}

Since the CGC- and TC-based variants of \textsc{OptiMetal2B} exhibit nearly identical scaling behavior in the 1D NSLs, we focus the 2D NSL analysis on the TC-based architectures of \textsc{OptiMetal2B} and \textsc{OptiMetal3B}.
This choice isolates the effect of increasing the body order from two to three.

To determine the functional form of the 2D NSL maps, we compared two candidate formulations using the AICc: a smoothly interpolating model inspired by Kaplan et al.~\cite{Kaplan2020} and an additive formulation following Hoffmann et al.~\cite{Hoffmann2022} (see Eqs.~(\ref{eq:kaplan_map}, \ref{eq:hoffmann_map}) in the Methods section).
The Kaplan-type formulation consistently yielded the lowest AICc and was therefore selected to describe $L(D,N)$.
In contrast to the 1D parameter-scaling fits, an irreducible loss term $L_\infty$ was no longer required, as both formulations naturally encode saturation at small $D$ and large $N$.

The resulting Kaplan-type 2D NSL maps based on Eq.~(\ref{eq:kaplan_map}) are shown in Fig.~\ref{fig:3dscaling}.
The top row shows $L(N)$ at fixed dataset sizes $D$, the middle row shows $L(D)$ at fixed parameter counts $N$, and the bottom row displays the full 2D landscape. 
The corresponding fit parameters are reported in Tab.~\ref{tab:scaling_maps}.

Looking at the top row of Fig.~\ref{fig:3dscaling}, we observe that, for both architectures, the performance improvement with increasing $N$ slows down once $N \gtrsim 10\,\mathrm{M}$, a trend that appears to be largely independent of dataset size.
However, the extent to which increasing $N$ improves performance prior to saturation depends strongly on the amount of training data: small datasets exhibit weak parameter scaling, whereas larger datasets benefit substantially more from increased model capacity before saturating.
In line with the 1D NSLs, both \textsc{OptiMetal2B} and \textsc{OptiMetal3B} show nearly identical parameter-scaling exponents ($\alpha_N=0.30$ and $0.33$), indicating that increasing body order does not significantly affect parameter scaling.

Turning to data scaling, we observe a pronounced dependence on model capacity, as seen in the middle row of Fig.~\ref{fig:3dscaling}.
For the smallest models, $L(D)$ initially follows a simple power law in the small-data regime but begins to saturate at around $D=40\,\mathrm{k}$, reflecting the limited capacity of the model in the high-data, low-parameter limit.
In the intermediate-parameter regime ($N \approx 1\,\mathrm{M}$), this bottleneck is lifted: both architectures exhibit a simple power-law data scaling behavior, indicating that model capacity and available data are well matched.
As the parameter count $N$ increases further, the data-scaling slope steepens, and for sufficiently overparameterized models, BNSLs re-emerge.
This effect is more pronounced for \textsc{OptiMetal3B} than for \textsc{OptiMetal2B}, consistent with the larger separation between its fitted exponents ($\alpha_{D,1}=0.16$ and $\alpha_{D,2}=0.46$) compared to \textsc{OptiMetal2B} ($\alpha_{D,1}=0.20$ and $\alpha_{D,2}=0.37$).
In contrast to parameter scaling, whose exponent remains nearly architecture-invariant, the data-scaling exponents differ substantially. 
This demonstrates that, while increasing body order improves data scaling, it does not effectively improve parameter scaling.
However, note that this effect only becomes apparent in the full 2D NSL analysis.

\begin{table}
    \centering
    \caption{
    \textbf{Fit parameters of the 2D NSL maps following Eq.~(\ref{eq:kaplan_map}) obtained using the AICc-based selection procedure described in the Methods section.}
    The reported parameters are obtained from nonlinear least squares fits to validation losses averaged over three random model initializations.
    The functional form in Eq.~(\ref{eq:kaplan_map}) combines the 1D data- and parameter-scaling laws summarized in Tab.~\ref{tab:scaling_laws}, while omitting the irreducible loss term $L_\infty$ in the parameter NSL, as explained in the Methods section.
    }
    \vspace*{2mm}
    \begin{tabular}{c@{\hspace{1em}}c@{\hspace{1em}}c@{\hspace{1em}}c@{\hspace{1em}}c@{\hspace{1em}}c@{\hspace{1em}}c@{\hspace{1em}}}
    \toprule
    Model & $\alpha_N$ & $\alpha_{D,1}$ & $\alpha_{D,2}$ & $N_0$ & $D_c$\\
    \midrule
    \textsc{OptiMetal2B} (TC) & $0.30$ & $0.20$ & $0.37$ & $10^{\,5.36}$ & $10^{\,4.51}$ \\
    \textsc{OptiMetal3B} (TC) & $0.33$ & $0.16$ & $0.46$ & $10^{\,5.14}$ & $10^{\,4.47}$ \\
    \bottomrule
    \end{tabular}
    \label{tab:scaling_maps}
\end{table}

Together, the 1D and 2D NSL analyses demonstrate that the broken data scaling observed in the 1D setting persists when the dataset size and parameter count are varied together, suggesting genuine multi-regime learning behavior.
Simple power-law scaling emerges when the size of the dataset and the capacity of the model are well matched. 
However, the global scaling behavior of invariant GNNs for optical properties is best described by a combination of smoothly broken data scaling and saturating parameter scaling, with the sharpness of the data-scaling transition depending on the body order of the architecture.

This behavior closely parallels observations in other areas of ML, where deviations from single power-law scaling have been associated with transitions between data-limited and capacity-limited regimes \cite{Hestness2017, Caballero2022}. 
Going beyond prior studies \cite{Hestness2017, Caballero2022, Frey2023}, we present evidence that these transitions persist when the dataset size and model capacity vary jointly. 
These findings underscore the importance of 2D NSL analyses for scientific ML tasks operating in the low-data, mid-capacity regime characteristic of materials science.

\section*{Discussion}

In this study, we examine neural scaling behavior for a high-dimensional spectroscopic learning task in materials science using over 200,000 ab initio dielectric functions of metals and invariant GNNs of controlled body order.

Our key finding is the emergence of BNSLs in data scaling. 
Rather than following a simple power law, model performance exhibits distinct scaling regimes, with a pronounced steepening of the data-scaling slope above a characteristic dataset size. 
In contrast, parameter scaling follows a saturating power law, indicating that the benefits of increasing model size diminish rapidly in the data-scarce regime of materials science.
By comparing invariant GNNs that differ only in their effective body order and message-passing formulation, we further isolate the influence of model architecture on scaling behavior. 
Variations in message passing seem to have little influence on the observed scaling trends, whereas increasing the body order improves data efficiency while leaving parameter scaling largely unchanged. 
Notably, the body-order dependence of the NSLs becomes apparent only in 2D NSL analyses that jointly consider dataset size and model capacity.

These findings are particularly relevant for materials science, where training data are scarce and costly to generate, yet model sizes remain modest compared to those in other ML domains.
Although increasing model capacity can provide rapid performance gains for a fixed dataset, this strategy may be impractical for many applications due to increased training cost and slower inference, especially for MLIPs \cite{Frank2024, Wood2025, Riebesell2025}.
Therefore, looking ahead, data availability and generation are likely to remain the primary bottlenecks for improving the predictive performance of ML models in materials science, motivating strategies that improve data efficiency.
In this context, two particularly active and competitive areas of research stand out.
One is the development of architectures that more explicitly encode physical symmetries, such as equivariant neural networks \cite{Batzner2022, Batatia2025}, which have been shown to substantially improve data efficiency, though at the cost of increased model complexity \cite{Ngo2025}.
The other is the strategic use of multi-fidelity and transfer learning strategies, which build upon models trained on large, inexpensive, lower-fidelity datasets to reduce the amount of expensive, high-fidelity data required for accurate predictions \cite{Zhang2024, Grunert2025, Baum2025}.
Such approaches are particularly attractive in domains where high-fidelity observables are computationally or experimentally expensive to obtain, such as optical spectra.

\section*{Methods}

\subsection{Dielectric functions of metals}

The dielectric linear response of a homogeneous solid in the optical limit can be expressed via the frequency-dependent dielectric tensor $\varepsilon_{\alpha\beta}(\omega)$. 
Within the IPA, and neglecting microscopic local-field effects \cite{Marini2001}, the total response, $\varepsilon_{\alpha\beta}$, is usually separated into interband, $\varepsilon_{\mathrm{inter}, \alpha\beta}(\omega)$, and intraband, $\varepsilon_{\mathrm{intra}, \alpha\beta}(\omega)$, contributions.
\begin{equation}
    \varepsilon_{\alpha\beta}(\omega) = \delta_{\alpha\beta} + \varepsilon_{\mathrm{inter}, \alpha\beta}(\omega) + \varepsilon_{\mathrm{intra}, \alpha\beta}(\omega)
\end{equation}
Within the above-mentioned limits and approximations, the interband term describes transitions between different bands and is given by \cite{Prandini2019Photo}
\begin{align}
    \label{eqn:inter}
    \varepsilon_{\mathrm{inter}, \alpha\beta}(\omega)\,=\,
    & -\frac{4\pi}{V} \sum_{\mathbf{k}}\sum_{\substack{n,n' \\ n \neq n'}} \frac{M^*_{nn'\mathbf{k}, \alpha} M_{nn'\mathbf{k},\beta}} {(E_{n'\mathbf{k}} - E_{n\mathbf{k}})^2} \nonumber \\
    &\times \frac{f_{n\mathbf{k}} - f_{n'\mathbf{k}}} {\omega - (E_{n'\mathbf{k}} - E_{n\mathbf{k}}) + i\gamma_\mathrm{inter}}
\end{align}
where $E_{n\mathbf{k}}$ are band energies and $f_{n\mathbf{k}}$ are Fermi-Dirac occupations of Bloch states $\Psi_{n\mathbf{k}}$ with band index $n$ at point $\mathbf{k}$ in the Brillouin zone.
$V$ is the unit-cell volume, and $\gamma_\mathrm{inter}$ is an ad hoc numerical broadening parameter.
$M_{nn'\mathbf{k}, \alpha}=\langle \Psi_{n'\mathbf{k}} | v_\alpha | \Psi_{n\mathbf{k}} \rangle$ are the matrix elements of the velocity operator $v_\alpha = -i [r_\alpha, H_\mathrm{KS}]$ along Cartesian direction $\alpha$, where $H_\mathrm{KS}$ is the Kohn-Sham (KS) Hamiltonian. 

The intraband contribution arises from carriers at the Fermi surface and is approximated by a Drude model \cite{Prandini2019Photo}
\begin{equation}
    \label{eqn:intra}
    \varepsilon_{\mathrm{intra}, \alpha\beta}(\omega)
    = -\frac{\omega_{\mathrm{D}, \alpha\beta}^2}{\omega\,(\omega+i\gamma_\mathrm{D})}
\end{equation}
with the Drude frequency tensor
\begin{equation}
    \label{eqn:drude}
    \omega_{\mathrm{D}, \alpha\beta}^2 = \frac{4\pi}{V}\sum_{\mathbf{k},n}
    \left(-\frac{\partial f_{n\mathbf{k}}}{\partial E_{n\mathbf{k}}}\,\right)\,M^*_{nn\mathbf{k},\alpha}\,M_{nn\mathbf{k},\beta} 
\end{equation}
where $\gamma_\mathrm{D}$ is the Drude damping, an ad hoc broadening parameter similar to $\gamma_\mathrm{inter}$. 
For semiconductors and insulators at $T=0$, the Drude frequency vanishes, i.e., $\omega_{\mathrm{D},\alpha\beta}=0$, leaving only the interband contribution.
Therefore, datasets of dielectric functions for semiconductors and insulators only contain the interband contribution $\varepsilon_{\mathrm{inter}, \alpha\beta}(\omega)$.

\subsection*{Ab initio calculations}

The DFT calculations were performed with the plane-wave code \textsc{Quantum ESPRESSO} \cite{Giannozzi2009all, Giannozzi2017all} using  optimized norm-conserving Vanderbilt pseudopotentials from the SG15 library \cite{Hamann2013} and PBE \cite{Marques2012} as exchange-correlation functional.
Dielectric and Drude frequency tensors in the IPA were calculated using the \textsc{SIMPLE} code \cite{Prandini2019}, which is part of the \textsc{Quantum ESPRESSO} distribution.
We selected the \textsc{SIMPLE} code because it implements the optimal basis (OB) method introduced by Shirley \cite{Shirley1996} for the periodic part of the Bloch wavefunctions, thus enabling their interpolation to arbitrary k-points \cite{Prendergast2009}.
Such interpolation is particularly valuable for optical calculations, as these usually require dense sampling of the Brillouin zone to converge.
This is especially true for metals, as their optical response depends sensitively on the precise electronic structure at the Fermi surface, requiring extremely fine k-point grids (see Supplementary Note 8 for an example).
Thus, the interpolation scheme implemented in \textsc{SIMPLE} allows for the efficient evaluation of optical properties on dense k-point grids.
However, even with this interpolation scheme, calculating and converging the dielectric functions for the 201,361 intermetallic compounds required approximately $2.3$ million CPU hours using the workflow described below.

The crystal structures for the calculations were taken from the \textsc{Alexandria} database \cite{Schmidt2023, Schmidt2024} (downloaded March 2024) of theoretically stable crystals and filtered by the following criteria: compounds containing only metallic elements from the first six rows of the periodic table (excluding lanthanides), ten or fewer atoms in the unit cell, a space group number greater than or equal to $75$, a distance from the convex hull of less than $250$~meV, and a non-magnetic ground state.
Following these criteria, we kept up to the five most stable crystal structures for each unique composition. 
After applying these filters, 205,224 intermetallic compounds remained, which were then reduced to their primitive standard structure using \textsc{Spglib} \cite{spglib}.

For every material, we performed the following three-step high-throughput workflow to calculate the diagonal components of the interband dielectric tensor, $\varepsilon_{\mathrm{inter},\alpha\alpha}(\omega)$, and the diagonal components of the Drude frequency tensor, $\omega_{\mathrm{D},\alpha\alpha}$.
All k-point grids ($k_x \times k_y \times k_z$) were generated from a structure-independent reciprocal density $n_\mathbf{k}$ per atom, as defined in \textsc{pymatgen} \cite{Ong2013}, using Monkhorst-Pack grids (i.e., the \texttt{automatic} k-point grid mode of \textsc{Quantum ESPRESSO}).
The number of subdivisions along each reciprocal-space direction was chosen to be even, such that any odd subdivision was increased to the next (even) integer.

First, we converged the DFT plane-wave cutoff energy and the k-point grid, employing the same strategy as in our previous $G_{0}W_{0}$-PPA benchmark workflow \cite{Grossmann2026}. 
Here, we adopted a looser convergence threshold of $1$~kcal/mol ($0.04$~eV per atom) for the total energy, rather than the $1$~meV per atom used in the benchmark setting, in order to reduce computational cost.
For all calculations, we used a "cold" smearing \cite{Marzari1999} of $300$~meV to improve convergence for the metallic systems at hand.
The resulting converged plane-wave cutoff energy and converged k-point grid were used in all subsequent steps, unless otherwise noted.

Second, following the converged ground state calculation, we performed a non-self-consistent DFT calculation on a uniform $2 \times 2 \times 2$ k-point grid without symmetry reduction, as required by \textsc{SIMPLE} for constructing the OB \cite{Prandini2019, Prandini2019Photo}.
Here, the number of bands included in the KS Hamiltonian was chosen such that all transitions up to $20$~eV above the Fermi energy were included.
Based on this calculation, we constructed the OB with the threshold for the Gram-Schmidt orthonormalization algorithm set to $10^{-2}$ a.u.~(input variable \texttt{s\_bands} in \textsc{SIMPLE} \cite{Prandini2019}).

Before describing the final step of the workflow, we briefly outline the settings used to compute optical properties in the IPA in \textsc{SIMPLE}.
During the band interpolation, \textsc{SIMPLE} first recalculates the Fermi energy, and then obtains the band occupations according to the Fermi-Dirac distribution at room temperature \cite{Prandini2019}. 
To ensure consistent Fermi energies, we used the same "cold" smearing \cite{Marzari1999} of $300$~meV ($\texttt{simpleip\_in\%fermi\_ngauss}=-1$, $\texttt{simpleip\_in\%fermi\_degauss}=0.02205$~Ry) as in the ground-state DFT calculations.
When evaluating the derivative of the Fermi-Dirac distribution in Eq.~(\ref{eqn:drude}), \textsc{SIMPLE} employs a Gaussian smearing to improve convergence of the Drude frequency with respect to the k-point grid \cite{Prandini2019}, which we set to $100$~meV ($\texttt{simpleip\_in\%drude\_degauss}=0.00735$~Ry).
Dielectric functions were evaluated on a uniform frequency grid of 2,001 points between $0$ and $20$~eV with $\gamma_\mathrm{inter}=300$~meV ($\texttt{simpleip\_in\%inter\_broadening}=0.02205$~Ry) and $\gamma_\mathrm{D}=100$~meV ($\texttt{simpleip\_in\%intra\_broadening}=0.00735$~Ry), cf.~Eq.~(\ref{eqn:inter}) and Eq.~(\ref{eqn:intra}).

In the third and final step, we converged $\varepsilon_{\mathrm{inter},\alpha\alpha}(\omega)$ and $\omega_{\mathrm{D},\alpha\alpha}$ with respect to the k-point grid.
We initialized the convergence loop with a k-point grid corresponding to $8\times n_\mathbf{k}$, effectively doubling the converged DFT k-point grid.
Since individual calculations are efficient due to interpolation, this dense starting point reduces the number of iterations required to achieve convergence.
Here, we increased $n_\mathbf{k}$ such that the number of subdivisions in at least two directions of the new grid exceeded those of the previous grid, i.e., $k_i^{n+1} > k^n_i$ for at least two directions $i = {x,y,z}$.
For each k-point grid, we computed $\varepsilon_{\mathrm{inter},\alpha\alpha}(\omega)$ and $\omega_{\mathrm{D},\alpha\alpha}$.
The procedure was repeated until (i) the change in the $\omega_{\mathrm{D},xx}$ between successive grids was below $100$~meV, and (ii) the similarity coefficient (SC) \cite{Grunert2024}
\begin{equation}\label{eqn:sc}
    \mathrm{SC}\left[\tilde{\varepsilon}(.);\varepsilon(.)\right] = 1 - \frac{\int\left\vert \varepsilon(\omega) -  \tilde{\varepsilon}(\omega)\right\vert\,d\omega}{\int \left\vert \varepsilon(\omega)\right\vert\,d\omega}
\end{equation}
between the rotationally invariant average of consecutive total dielectric tensors, i.e., $\overline{\varepsilon}(\omega)=\mathrm{Tr}[\varepsilon_{\alpha\beta}(\omega)]/3$, evaluated from $1$--$20$~eV, exceeded $0.98$.
In Eq.~(\ref{eqn:sc}), $\tilde{\varepsilon}(\omega)$ refers to the dielectric function computed on the current k-point grid, and $\varepsilon(\omega)$ refers to that computed on the preceding, less dense k-point grid.
The k-point grid convergence loop terminated once both criteria were satisfied or after 20 iterations. 
In the latter case, the material was discarded from the final dataset.

For each material, we stored the crystal structure, along with all information relating to the convergence iterations for the total energy and optical properties, the converged $\varepsilon_{\mathrm{inter},\alpha\alpha}(\omega)$, and $\omega_{\mathrm{D},\alpha\alpha}$ in the form of a \texttt{ComputedStructureEntry} from \textsc{pymatgen} \cite{Ong2013}, which is essentially a JSON file.

\subsection*{Machine learning}

All ML models were implemented using \textsc{PyTorch} and \textsc{PyTorch Geometric}.
For training and evaluating the models, the dataset was split by unique chemical composition \cite{Merchant2023, Grunert2024} into fixed training, validation, and test sets in an 80:10:10 ratio, each containing 160,728, 20,179, and 20,454 materials, respectively.
Supplementary Note 9 provides statistics on the prevalence of elements and the distribution of unit cell sizes across the training, validation, and test sets.
As stated at the beginning of the Results section, for each material we used rotationally invariant averages of the interband dielectric tensor, $\overline{\varepsilon}_\mathrm{inter}(\omega) = \mathrm{Tr}[\varepsilon_{\mathrm{inter},\alpha\beta}(\omega)]/3$, and Drude frequency tensor, $\overline{\omega}^{\,2}_\mathrm{D} = \mathrm{Tr}[\omega_{\mathrm{D},\alpha\beta}^{2}]/3$, as targets for our invariant GNNs. 
For each material, the primary graph was created by converting crystal structures into multigraphs \cite{Xie2018}.
Each node in the primary graph corresponds to an atom in the unit cell, and edges are created between nodes if the distance between the corresponding atoms is less than or equal to $r^\mathrm{2B}_{c}=5.5$~{\AA}, taking into account periodic boundary conditions.
The line graph, required for \textsc{OptiMetal3B}, was then obtained by connecting edges whose lengths are less than or equal to $r^\mathrm{3B}_{c}=4.0$~{\AA}, in line with \textsc{M3GNet} \cite{Chen2022} and \textsc{MatterSim} \cite{Yang2024}.
As a loss function for our multi-objective GNNs, we combine the mean absolute errors (MAE) 
\begin{equation}
    \mathrm{MAE}\left[Y; \tilde{Y}\right] = \frac{1}{N}\sum_{i=1}^N \left\vert Y_i - \tilde{Y}_i \right\vert 
\end{equation}
of $\overline{\varepsilon}_\mathrm{inter}(\omega)$ and $\overline{\omega}_\mathrm{D}$
\begin{equation}   
L = \mathrm{MAE}\left[\overline{\varepsilon}^\mathrm{ML}_\mathrm{inter}(\omega); \overline{\varepsilon}^\mathrm{DFT}_\mathrm{inter}(\omega)\right] + \mathrm{MAE}\left[\overline{\omega}^\mathrm{ML}_\mathrm{D}; \overline{\omega}^\mathrm{DFT}_\mathrm{D}\right]
\end{equation}
following our previous finding that MAE-based training performs better than MSE-based training for optical spectra \cite{Grunert2024}.
Throughout all models, the rectified linear unit (ReLU) was used as the activation function in all layers involving nonlinearities.

All models were trained for 500 epochs with \textsc{AdamW} \cite{Loshchilov2019}, employing a cosine annealing learning rate schedule. 
Each training run began with a 5-epoch warm-up phase, during which the learning rate increased linearly from zero to the initial maximum learning rate $\eta_\mathrm{max}$ of the cosine annealing schedule.
Note that the warm-up epochs were included in the total training epoch count. 
After the warm-up phase, the learning rate gradually decayed to zero over the remaining epochs using a half-cosine cycle.
We applied gradient clipping with a norm threshold of $100$ and used a fixed batch size of $256$ for all training runs.
Each model was trained with three different random initializations to ensure robust performance metrics.
All training and inference runs were performed on \textsc{NVIDIA} A100 GPUs in mixed-precision mode using \texttt{bfloat16} to reduce memory usage and accelerate computation.

All architectural choices and optimizer hyperparameters were fixed to the values identified during the architecture optimization described in Supplementary Note 4, which was performed with $d_\mathrm{h}=256$ using the 20,000-sample training subset.
However, we found that training became unstable for TC-based models at hidden dimensions $d_\mathrm{h} > 256$ when using the maximum learning rate $\eta_\mathrm{max}$ optimized for $d_\mathrm{h} = 256$.
To address this issue, we scaled $\eta_\mathrm{max}$ proportionally to the number of trainable parameters, $N$, ensuring that the resulting scaling laws reflect dependence solely on $N$ rather than on width-specific learning-rate re-optimization.
Based on additional experiments reported in Supplementary Note 10, we found that a learning-rate scaling exponent of one provided the best overall stability and performance.

A total of 8,100 GPU hours were used to train all models for architecture optimization and scaling law analysis.

\subsection{Functional forms of neural scaling laws}

Guided by previous studies on NSLs \cite{Hestness2017, Kaplan2020, Hoffmann2022, Caballero2022, Frey2023, Ngo2025}, we evaluated four candidate functional forms for the 1D NSL analysis:
\begin{align}
    L(X) &= \left(\frac{X_0}{X}\right)^{\alpha} \label{eq:pl}\\
    L(X) &= L_\infty + \left(\frac{X_0}{X}\right)^{\alpha} \label{eq:pl_floor}\\
    L(X) &= 
    \left(\frac{X_c}{X}\right)^{\alpha_2}
    \left[1 + \left(\frac{X_c}{X}\right)\right]^{\alpha_1 - \alpha_2} \label{eq:broken}\\
    L(X) &= 
    A\left(\frac{X_c}{X}\right)^{\alpha_2}
    \left[1 + \left(\frac{X_c}{X}\right)\right]^{\alpha_1 - \alpha_2} \label{eq:broken_amp}
\end{align}
Here, $X\in\{D,N\}$ denotes either the dataset size ($D$) or the number of trainable parameters ($N$).  
In the simple NSLs in Eqs.~(\ref{eq:pl}, \ref{eq:pl_floor}), $X_0$ sets the point where $L(X_0)=1$, while in the BNSLs in Eqs.~(\ref{eq:broken}, \ref{eq:broken_amp}), $X_c$ marks the crossover where the scaling exponent changes.
The parameters $\alpha$, $\alpha_1$, and $\alpha_2$ control the decay of the loss with increasing $X$, while $L_\infty$ represents a possible irreducible loss and $A$ an optional multiplicative amplitude.
In learning settings based on ab initio data, we generally expect that, in the limit of sufficiently large $D$ and $N$, $L_\infty = 0$, since no residual entropy or irreducible uncertainty is expected.
This is in contrast to natural-language models, where an asymptotic loss floor is unavoidable.
Each of the four functions was fitted to the validation loss, averaged over three random model initializations, using nonlinear least-squares regression.

To compare the descriptive quality of these functions, we used the Akaike information criterion (AIC), which balances how well a function fits the data with the number of parameters used for the fit \cite{Burnham2002}.  
For least-squares fitting, the AIC is given by
\begin{equation}
    \mathrm{AIC}
    = 2k
    - 2\ln\!
    \left(
    \frac{1}{n}
    \sum_{i=1}^{n}
    \left[L(X_i) - L_{\mathrm{val}}(X_i)\right]^{\,2}
    \right)
    \label{eq:AIC}
\vspace{0.2cm} 
\end{equation}
where $k$ is the number of fit parameters and $n$ is the number of data points used for the fit.
Since the 1D NSLs fits are based on only seven data points, we applied the finite-sample correction to obtain the Akaike information criterion corrected for small sample sizes (AICc)
\begin{equation}
    \mathrm{AICc}
    = \mathrm{AIC}
    + \frac{2k^2+2k}{n - k - 1}
    \label{eq:AICc}
\end{equation}
which prevents overly flexible models from being favored when $n$ is small relative to $k$ \cite{Burnham2002}.  
For each model and 1D NSL, the function with the lowest AICc value was chosen as the preferred model.
The corresponding AICc values can be found in Supplementary Note 7.

After determining the 1D NSL forms that best describe the data and parameter scaling---broken data scaling following Eq.~(\ref{eq:broken}) and saturated parameter scaling following Eq.~(\ref{eq:pl_floor})---we constructed 2D NSL maps $L(D,N)$ by combining their respective dependencies.  
Inspired by the work of Kaplan et al.~\cite{Kaplan2020}, we tested the smoothly interpolating form
\begin{widetext}
    \begin{equation}
        \label{eq:kaplan_map}
        L(D,N)
        =
        \left[\,
            \left(\frac{N_0}{N}\right)^{\tfrac{\alpha_N}{\alpha_{D,2}}}
            +
            \left(\frac{D_c}{D}\right)
            \left(1 + \frac{D_c}{D}\right)^{\left(\tfrac{\alpha_{D,1}}{\alpha_{D,2}} - 1\right)}
        \,\right]^{\,\alpha_{D,2}}
    \end{equation}
\end{widetext}
In contrast to the 1D NSL fit for parameter scaling, we found that data can already be explained well without an irreducible loss, $L_\infty$, as saturation at low $D$, with respect to $N$, is already included in Eq.~(\ref{eq:kaplan_map}).
As an alternative formulation, following Hoffmann et al.~\cite{Hoffmann2022}, we also tested the additive description
\begin{widetext}
    \begin{equation}
        \label{eq:hoffmann_map}
        L(D,N)
        =
        \left(\frac{N_0}{N}\right)^{\alpha_N}
        +
        \left(\frac{D_c}{D}\right)^{\alpha_{D,2}}
        \left(1 + \frac{D_c}{D}\right)^{\alpha_{D,1}-\alpha_{D,2}}
    \end{equation}
\end{widetext}
where again the irreducible loss $L_\infty$ was not required as the fit found it to be zero.
Both 2D NSL maps were fitted using nonlinear least-squares regression, and the functional form with the lowest AICc value was chosen as the preferred representation of $L(D,N)$.  
The corresponding AICc values are provided in Supplementary Note 7.

\section*{Data Availability}

The dataset containing the crystal structures, interband dielectric functions, and Drude frequencies of 201,361 intermetallic compounds is distributed across two repositories due to its size: \url{https://doi.org/10.6084/m9.figshare.31111798} and \url{https://doi.org/10.6084/m9.figshare.31112491}.
The model weights for the architectures evaluated in Supplementary Note 5 are available at \url{https://github.com/MaxGrossmann/optimetal}.
TensorBoard log files of all training runs, along with the training, validation, and test splits of the dataset in compressed HDF5 format, and the derived graph representations, are available at \url{https://doi.org/10.6084/m9.figshare.31112554}.

\section*{Code Availability}

The third-party electronic structure codes \textsc{Quantum ESPRESSO} and \textsc{SIMPLE} are available at \url{https://quantum-espresso.org}. 
The high-throughput workflow used in this study, including the filtered input structures derived from the \textsc{Alexandria} database, is available at \url{https://github.com/MaxGrossmann/ht_metals}.
All code and scripts used to implement, train, evaluate, and optimize the architecture of the machine learning models, as well as those used to obtain the neural scaling laws, are available at \url{https://github.com/MaxGrossmann/optimetal}.

\section*{Acknowledgments}

We thank the staff of the Compute Center of the Technische Universität Ilmenau, especially Mr.~Henning~Schwanbeck for providing an excellent research environment. 
M.~Großmann thanks C.~Dreßler, J.~H\"anseroth, and A.~Fl\"ototto for their valuable feedback on the first draft of the manuscript.
This work is supported by the Deutsche Forschungsgemeinschaft DFG (project 537033066) and the Carl-Zeiss-Stiftung (funding code: P2023-02-008).

\section*{Competing interests}

The authors declare no competing interests.

\section*{Author contributions}
M.G.~and M.G.~conceived the idea; M.~Großmann~wrote the high-throughput workflow, performed all calculations, analyzed the data, trained and optimized the ML models, performed the scaling law analysis, visualized all results, and wrote the first draft of the manuscript.
E.R.~supervised the work and contributed to the interpretations; all authors revised and approved the manuscript.

\bibliography{literature}

\end{document}


\title{Supplementary Information for "Broken neural scaling laws in materials science"}

\author{Max Großmann}
\email{max.grossmann@tu-ilmenau.de}
\affiliation{Institute of Physics and Institute of Micro- and Nanotechnologies, Technische Universit\"at Ilmenau, 98693 Ilmenau, Germany}

\author{Malte Grunert}
\affiliation{Institute of Physics and Institute of Micro- and Nanotechnologies, Technische Universit\"at Ilmenau, 98693 Ilmenau, Germany}

\author{Erich Runge}
\affiliation{Institute of Physics and Institute of Micro- and Nanotechnologies, Technische Universit\"at Ilmenau, 98693 Ilmenau, Germany}

\date{\today}

\maketitle

For definitions of abbreviations, please refer to the main text, where all abbreviations are defined in detail. 
Abbreviations not introduced in the main text are defined here.

\tableofcontents
\newpage
\clearpage

\section*{Supplementary Note 1: Ab initio dataset composition}

\begin{figure}[ht]
    \centering
    \includegraphics{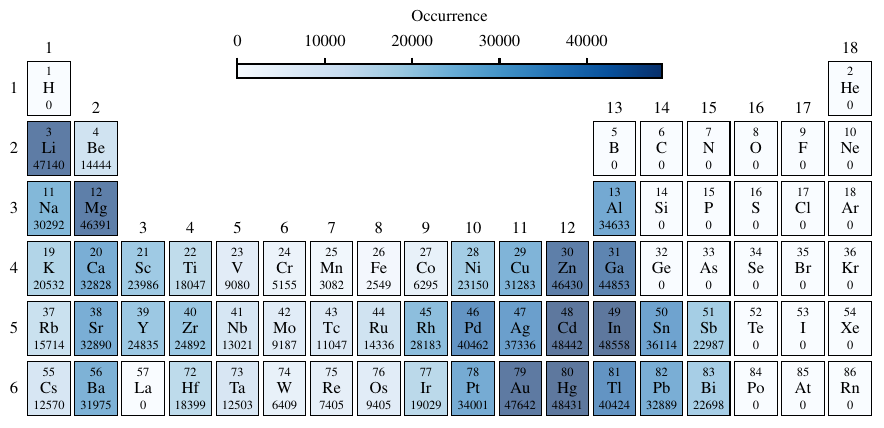}
    \caption{
    \textbf{Periodic table illustrating the elemental distribution in the full ab initio dataset produced in this study.}
    Colors indicate the number of occurrences of each element in the dataset, with exact counts shown below the respective symbols.
    The lanthanides and the seventh period are omitted, as there are no elements from these groups in the dataset.
    }
    \label{fig:pse}
\end{figure}

\begin{figure}[ht]
    \centering
    \includegraphics{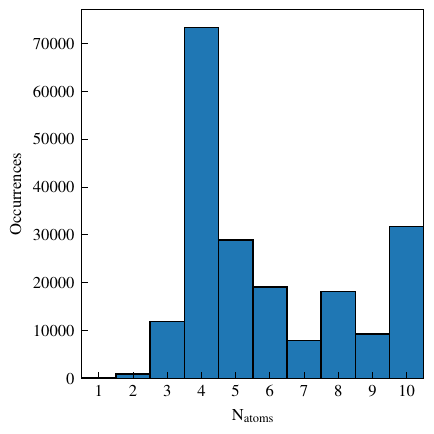}
    \caption{
    \textbf{Distribution of the number of atoms per unit cell for all compounds in the full ab initio dataset produced in this study.}
    }
    \label{fig:nsites}
\end{figure}

\newpage
\clearpage

\section*{Supplementary Note 2: Validation against experimental dielectric functions}

\begin{table}[ht]
    \centering
    \caption{
    \textbf{List of elemental metals used to validate the quality of the calculation methodology and high-throughput ab initio workflow used in this study.}
    The table contains the \textsc{Materials Project} identifiers (MP-ID) \cite{Jain2013,Ong2015} of the used crystal structures, as well as the corresponding experimental references (Ref.) with which the calculated dielectric functions were compared.
    }
    \vspace*{2mm}
    \begin{tabular}{ll@{\hspace{1em}}r
    }
    \toprule
    Material~ & MP-ID & Ref.\\
    \midrule
    Ag & mp-124 & \cite{Babar2015} \\
    Al & mp-134 & \cite{Cheng2016} \\
    Au & mp-81  & \cite{Babar2015} \\
    Be & mp-87  & \cite{Palik1998} \\
    Ca & mp-45  & \cite{Mathewson1971} \\
    Cs & mp-1   & \cite{Smith1970} \\
    Cu & mp-30  & \cite{Babar2015} \\
    In & mp-85  & \cite{Mathewson1971} \\
    Ir & mp-101 & \cite{Schmitt2022} \\
    K  & mp-58  & \cite{Smith1969} \\
    Li & mp-135 & \cite{Inagaki1976} \\
    Mg & mp-153 & \cite{Palm2018} \\
    Mo & mp-129 & \cite{Palik1998} \\
    Na & mp-127 & \cite{Inagaki1976} \\
    Nb & mp-75  & \cite{Weaver1973} \\
    Os & mp-49  & \cite{Palik1998} \\
    Pd & mp-2   & \cite{Johnson1974} \\
    Pt & mp-126 & \cite{Palik1998} \\
    Rb & mp-70  & \cite{Smith1970} \\
    Re & mp-8   & \cite{Palik1998} \\
    Rh & mp-74  & \cite{Weaver1977} \\
    Ru & mp-33  & \cite{Palik1998} \\
    Ta & mp-50  & \cite{Palik1998} \\
    Ti & mp-46  & \cite{Johnson1974} \\
    V & mp-146  & \cite{Johnson1974} \\
    W & mp-91   & \cite{Weaver1975} \\
    Zr & mp-131 & \cite{Querry1987} \\
    \bottomrule
    \end{tabular}
    \label{tab:simple_metals}
\end{table}

To validate the calculation methodology and high-throughput workflow used in this study, we benchmarked the calculated dielectric functions against experimental reference data for 27 elemental metals, which is a representative, albeit limited, set due to availability of experimental data.
Crystal structures were taken from the \textsc{Materials Project}, and the corresponding identifiers, along with references for the corresponding measured dielectric functions, are listed in Supplementary Tab.~\ref{tab:simple_metals}.

As shown in Supplementary Figs.~\ref{fig:exp1}--\ref{fig:exp3}, we observe good overall agreement between the ab initio and experimental spectra, consistent with previous results by Prandini et al.~\cite{Prandini2019Photo}.
Nevertheless, systematic deviations are apparent in the absorption features associated with transitions involving occupied $d$ states, such as the $d^{\,10}s^{1}\!\rightarrow d^{\,9}s^{2}$ transitions in Ag and Au.
These deviations originate from the well-known tendency of semi-local functionals, such as PBE, to position occupied $d$ bands too close to the Fermi level.
As a consequence, transitions involving these states are underestimated in energy, leading to a redshift of the corresponding absorption features relative to experiment.
This behavior reflects the incomplete representation of electronic correlation by Kohn-Sham bands and the resulting consequences for describing $d$ states of transition metals---a topic extensively discussed in the literature \cite{Fulde1995,Runge1996,Pavarini2012}.

More accurate electronic structure methods can substantially reduce these deficiencies.
To illustrate this, we calculated the IPA dielectric function of Ag and Au starting from QS$GW$ \cite{Kotani2007} electronic structures and included the resulting spectra in Supplementary Fig.~\ref{fig:exp1} (blue lines).
These calculations were performed with the all-electron LMTO code \textsc{Questaal} (version 7.14.1) \cite{Pashov2020} using the workflow introduced in Ref.~\cite{Großmann2026} with minor adjustments.
Specifically, the self energy was converged such that the energies of the valence-band maximum and conduction-band minimum at the $\Gamma$ point were converged to within 25~meV, following the convergence criterion suggested in Ref.~\cite{Grossmann2024}.
For the interband dielectric function, a $24\times24\times24$ ($28\times28\times28$) k-point grid was used for Ag (Au), while a $48\times48\times48$ ($56\times56\times56$) k-point grid was used to calculate the Drude frequency.
Here, Brillouin zone integrations were performed using the tetrahedron method \cite{Bloechl1994}.

Although performing QS$GW$ calculations on the entire dataset would be too costly, doing so on a smaller scale can still be beneficial, as the results can be used for transfer learning \cite{Grunert2025}, providing a natural route for further improving prediction quality in future work.

\begin{figure*}[ht]
    \centering
    \includegraphics{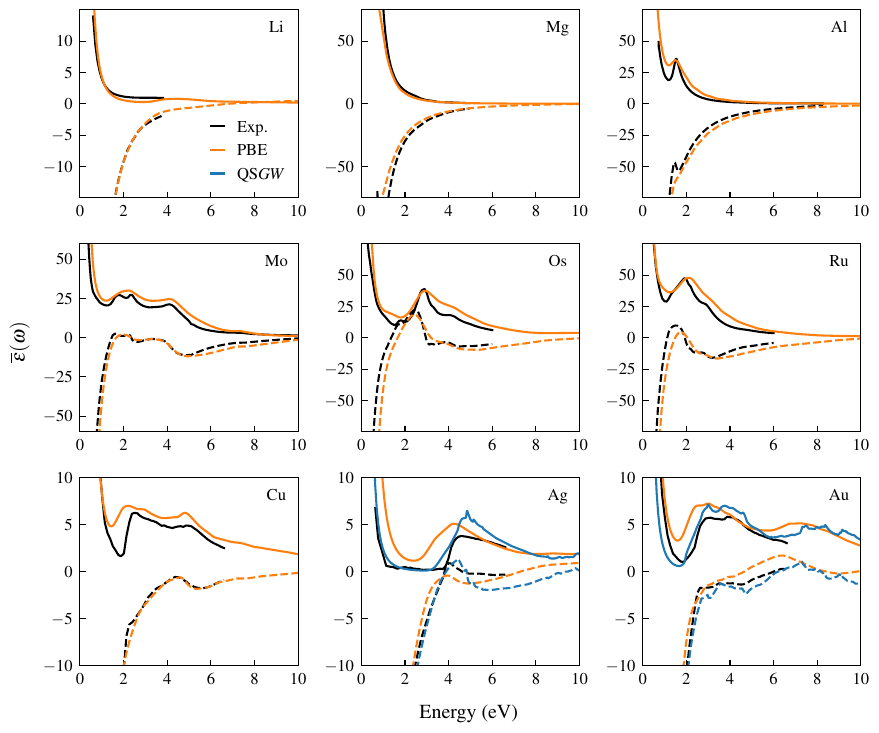}
    \caption{
    \textbf{Comparison of calculated and experimental dielectric functions of elemental metals.}
    The complex dielectric functions (solid and dashed lines: imaginary and real part of $\overline{\varepsilon}(\omega)$, respectively) of elemental metals were calculated using crystal structures obtained from the \textsc{Materials Project} \cite{Jain2013,Ong2015}.
    The black lines show the experimental reference data, while the orange lines illustrate the results of the high-throughput ab initio workflow described in the Methods section of the main text.
    For Ag and Au, additional QS$GW$ calculations are shown, illustrating that optical transitions involving occupied $d$ states are not accurately captured at the DFT level, leading to systematic deviations in the interband region.
    Details of the QS$GW$ calculations are provided in this Supplementary Note.
    The \textsc{Materials Project} identifiers \cite{Jain2013,Ong2015} for the crystal structures and experimental references are listed in Supplementary Tab.~\ref{tab:simple_metals}.
    }
    \label{fig:exp1}
\end{figure*}

\begin{figure*}[ht]
    \centering
    \includegraphics{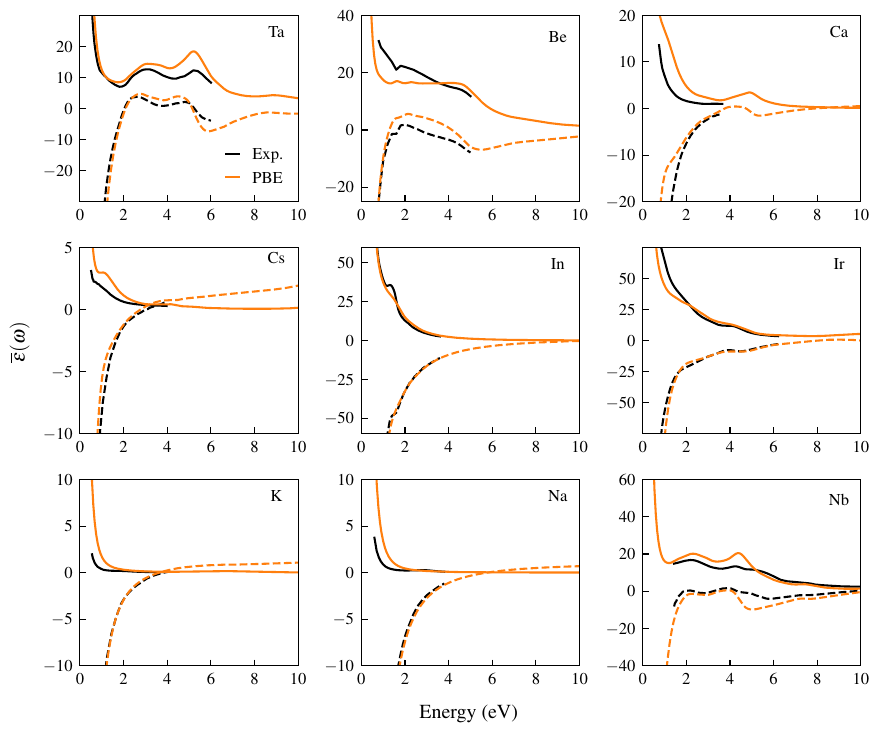}
    \caption{
    \textbf{Comparison of calculated and experimental dielectric functions of elemental metals.}
    Continuation of Supplementary~Fig.~\ref{fig:exp1}.
    }
    \label{fig:exp2}
\end{figure*}

\begin{figure*}[ht]
    \centering
    \includegraphics{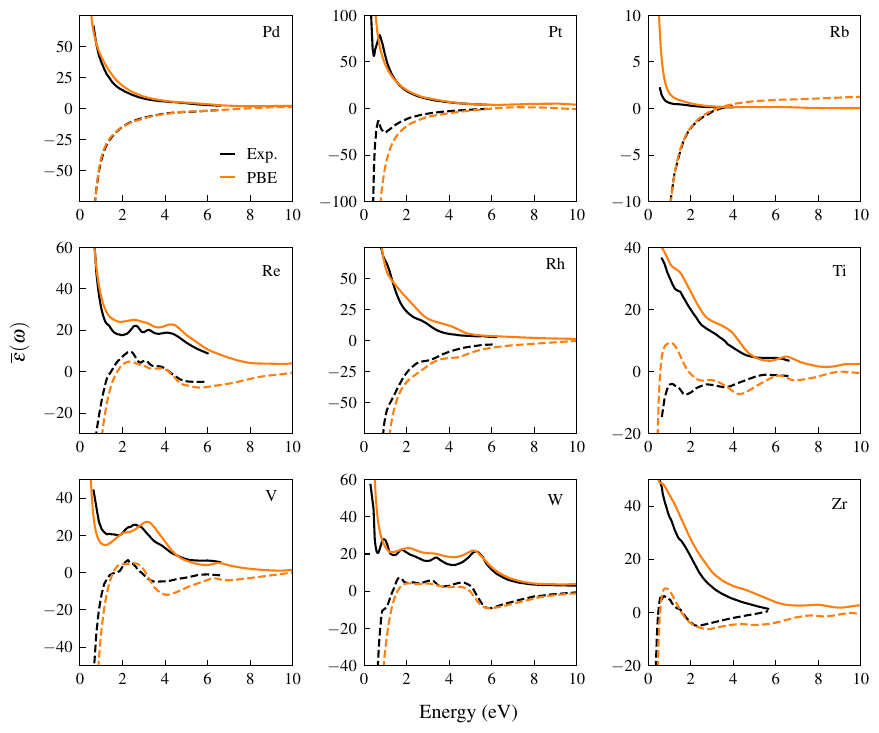}
    \caption{
    \textbf{Comparison of calculated and experimental dielectric functions of elemental metals.}
    Continuation of Supplementary~Fig.~\ref{fig:exp1}.
    }
    \label{fig:exp3}
\end{figure*}

\newpage
\clearpage

\section*{Supplementary Note 3: Model architecture}

\begin{figure*}[ht]
    \centering
    \includegraphics{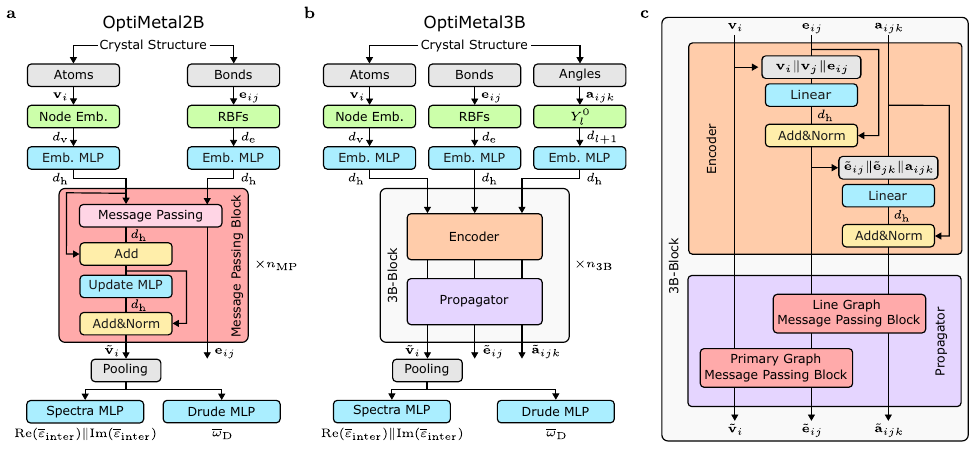}
    \caption{
    \textbf{Sketch of the employed GNN architectures and the three-body message passing block used in this work.}
    \textbf{a}, Architecture of \textsc{OptiMetal2B}.
    \textbf{b}, Architecture of \textsc{OptiMetal3B}.
    \textbf{c}, Structure of the three-body block (\textsc{3B-Block}) used in \textsc{OptiMetal3B} in \textbf{b}.
    The following abbreviations are used in the schematic: multilayer perceptron (MLP), radial basis functions (RBFs), embedding (Emb.), and spherical harmonics $Y_l^0$.
    Latent dimensions are denoted by the variable $d$, with subscripts indicating their role: $d_\mathrm{v}$ for node embeddings, $d_\mathrm{e}$ for edge embeddings, $d_{l+1}$ for angle embeddings, and $d_\mathrm{h}$ for the general hidden dimension.
    Concatenation of latent vectors is represented by the $\|$ operator.
    Blocks correspond to the same underlying layers and are color-coded consistently across all panels, e.g., all embedding layers are light green and all MLPs are light blue.
    Each panel and the function of each block are described in detail in this Supplementary Note.
    }
    \label{fig:architecture}
\end{figure*}

This Supplementary Note describes the design of the two GNN architectures used in this work, \textsc{OptiMetal2B} and \textsc{OptiMetal3B}.
The overall architecture of \textsc{OptiMetal2B} and its three-body extension, \textsc{OptiMetal3B}, is sketched in Supplementary Fig.~\ref{fig:architecture}.
Starting from a crystal structure, it is converted into a multigraph \cite{Xie2018}, where each atom is encoded as a node and edges represent bonds between atoms (cf.~Methods in the main text).
This primary graph of atoms ($\mathbf{v}_i$) and bonds ($\mathbf{e}_{ij}$), highlighted by the gray boxes in Supplementary Fig.~\ref{fig:architecture}a, serves as the input to \textsc{OptiMetal2B}.
\textsc{OptiMetal3B} extends this representation by introducing angles between bonds and constructing a so-called line graph \cite{Choudhary2021, Ruff2024}, similar to \textsc{M3GNet} \cite{Chen2022} and \textsc{MatterSim} \cite{Yang2024}.
The line graph is derived from the primary graph, such that each node in the line graph corresponds to an edge in the primary graph.
Consequently, two nodes (bonds) in the line graph are connected if their corresponding edges in the primary graph share a common node (atom), and edges in the line graph therefore correspond to angles.
The input to \textsc{OptiMetal3B} is thus composed of the primary graph and its line graph, represented by atoms ($\mathbf{v}_i$), bonds ($\mathbf{e}_{ij}$), and angles ($\mathbf{a}_{ijk}$) shown in gray in Supplementary Fig.~\ref{fig:architecture}b.

Now, we briefly outline the main components of both architectures.
Generic labels such as "embedding", "message passing", and "pooling" are used in Supplementary Fig.~\ref{fig:architecture} to emphasize the architectural structure rather than specific layer choices.
The specific layers are selected later through an optimization procedure, with choices restricted to the options listed in the following.

In both \textsc{OptiMetal2B} and \textsc{OptiMetal3B}, node and edge features are embedded into a feature space.
The embedding of atomic species on nodes was done using either a concatenation of one-hot encodings of the group and period of each element (group-period embedding) \cite{Grunert2024}, or a learned embedding layer---analogous to the token embeddings used in LLMs.
Inspired by state-of-the-art interatomic potentials \cite{Fu2025}, Gaussian and Bessel radial basis functions (RBFs) (with and without multiplication by a smooth polynomial envelope function \cite{Gasteiger2020}) were considered as embeddings for bond lengths on edges.
In \textsc{OptiMetal3B}, angles were embedded as spherical harmonics, $Y_l^0$, following \textsc{M3GNet} \cite{Chen2022} and \textsc{MatterSim} \cite{Yang2024}, and no other embedding methods were explored.
After the node, edge, and angle embeddings are created, each is passed through an individual multilayer perceptron (MLP), which acts as a nonlinear projector and maps the embeddings into a latent space.
Here, each MLP consists of two hidden layers with a fixed hidden-layer width $d_\mathrm{h}$.
Note that the number of hidden layers was fixed in order to reduce the number of free architectural parameters.
Meanwhile, $d_\mathrm{h}$, which is also used in subsequent network blocks, defines the overall model width and thereby controls the number of trainable model parameters $N$.

In \textsc{OptiMetal2B}, after the initial embedding layers and MLPs, messages between nodes and edges are aggregated within the \textsc{Message Passing Block}.
This block consists of a standard message-passing layer---specifically \textsc{CGConv} (CGC) \cite{Xie2018}, \textsc{GATv2Conv} (GATC) \cite{Brody2021}, or \textsc{TransformerConv} (TC) \cite{Thekumparampil2018}---followed by a one-hidden-layer MLP whose output is merged with the residual connection through element-wise addition, and then normalized at the graph level (\textsc{Add\&Norm}).
Thus, the resulting structure, sketched in the red block in Supplementary Fig.~\ref{fig:architecture}a, mirrors the feed-forward sub-block of a modern Transformer---a design that has proven central to the success of LLMs \cite{Vaswani2017, Grattafiori2024}.
The width of the one-hidden-layer MLP was set to $w_\mathrm{MLP} = m_\mathrm{MLP} \times d_\mathrm{h}$, where the one-hidden-layer MLP width multiplier $m_\mathrm{MLP}$ is an architectural hyperparameter optimized during the architecture optimization described in Supplementary Note 4.
As shown in Supplementary Fig.~\ref{fig:mlp_addnorm}, using the baseline \textsc{OptiMetal2B} architecture (see Supplementary Note 4) as a reference, incorporating the residual MLP+\textsc{Add\&Norm} operation after each message-passing layer consistently reduces the validation loss by about 10\% across all tested message-passing depths.
This reduction exceeds the expected improvement from the associated increase in parameter count alone.

\begin{figure}[ht]
    \centering
    \includegraphics{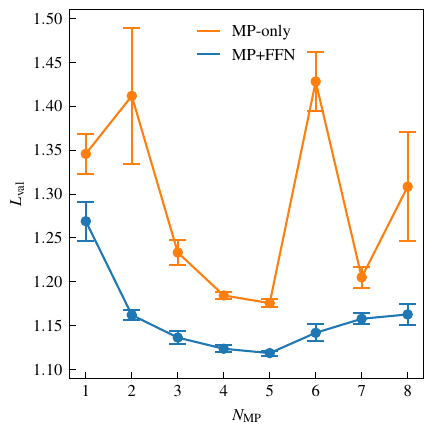}
    \caption{
    \textbf{Effect of incorporating a residual MLP+\textsc{Add\&Norm} operation after each message-passing layer, evaluated for the baseline \textsc{OptiMetal2B} architecture.}
    The validation loss, $L_\mathrm{val}$, is shown as a function of the number of message-passing layers, $N_\mathrm{MP}$, comparing standard message passing (MP-only) with a variant in which each message-passing layer is followed by a one-hidden-layer MLP with a residual connection and graph-level normalization (MP+MLP+\textsc{Add\&Norm}, denoted as MP+FNN in the legend).
    Dots represent $L_\mathrm{val}$ averaged over three random model initializations, and error bars indicate the standard deviation.
    All models were trained for $500$ epochs using the subset of the training set consisting of 20,000 materials (see Supplementary Note 6), with a fixed hidden dimension of $d_\mathrm{h}=256$.
    The optimizer hyperparameters were optimized for each model using the first stage of the workflow described in Supplementary Note 4.
    }
\label{fig:mlp_addnorm}
\end{figure}

In \textsc{OptiMetal3B}, message aggregation is more involved, since information from the primary and line graph needs to be coupled, as shown in Supplementary Fig.~\ref{fig:architecture}b and c.
This coupling is achieved through the \textsc{3B-Block}, consisting of an \textsc{Encoder} and a \textsc{Propagator} sub-block, as illustrated in Supplementary Fig.~\ref{fig:architecture}c.
In the \textsc{Encoder}, each bond and angle is enriched with information about its surrounding atomic environment.
Specifically, node features are concatenated with their adjacent edge features and passed through a linear transformation followed by an \textsc{Add\&Norm} operation, updating the edge representations.
The same procedure is then applied to pass information from the edges to the angles. 
This allows each bond (angle) to learn the chemical context of the atoms (bonds) it connects in the primary (line) graph.
The \textsc{Propagator} performs the inverse operation by first updating the bonds through a \textsc{Message Passing Block} on the line graph and then propagating information back to the nodes through a \textsc{Message Passing Block} on the primary graph.
This bidirectional exchange---from nodes to edges to angles and back---establishes a strong connection between the primary and line graphs, improving the model's ability to capture three-body interactions.

In line with the depth of the embedding MLPs, the \textsc{Message Passing Block} in \textsc{OptiMetal2B} and the \textsc{3B-Block} in \textsc{OptiMetal3B} were applied twice in all networks studied here to reduce the number of free architectural parameters.
Similarly, the dimension of the latent space for nodes and edges remained constant at $d_\mathrm{h}$ throughout all operations.
The primary and line graph in the \textsc{3B-Block} use the same \textsc{Message Passing Blocks}.

\begin{figure}[ht]
    \centering
    \includegraphics{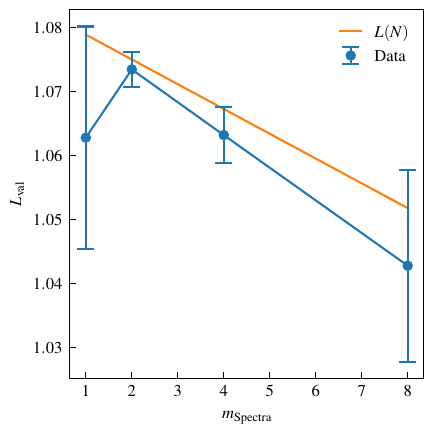}
    \caption{    
    \textbf{Effect of the \textsc{Spectra MLP} width multiplier on model performance, evaluated for the optimized TC-based \textsc{OptiMetal2B} architecture.}
    The validation loss $L_\mathrm{val}$ is shown as a function of the width multiplier of the \textsc{Spectra MLP}, $m_\mathrm{Spectra}$.
    All models were trained for $500$ epochs using the subset of the training set consisting of 20,000 materials (see Supplementary Note 6), with a fixed hidden dimension of $d_\mathrm{h}=256$.
    Thus, increasing $m_\mathrm{Spectra}$ only increases the number of trainable parameters in the \textsc{Spectra MLP}.
    Dots represent $L_\mathrm{val}$ averaged over three random model initializations, and error bars indicate the standard deviation.
    The solid orange line shows the parameter-scaling law $L(N)$ obtained from the 1D NSL analysis (see main text), indicating that variations in $m_\mathrm{Spectra}$ affect performance primarily through the associated change in model size.
    }
    \label{fig:spectra_dim}
\end{figure}

In both architectures, after message aggregation, the node features are pooled into a global latent vector using either mean, scalar/vector attention-based \cite{Li2019, Grunert2024}, or \textsc{Set2Set} \cite{Vinyals2015} pooling, yielding a compact representation of each material in the latent materials space \cite{Grunert2025Online}.
This global representation is then processed by two independent output MLPs (see Supplementary Fig.~\ref{fig:architecture}a and b): a \textsc{Spectra MLP} that predicts the interband dielectric function, $\mathrm{Re}\left(\overline{\varepsilon}_\mathrm{inter}\right)\|\,\mathrm{Im}\left(\overline{\varepsilon}_\mathrm{inter}\right)$, and a \textsc{Drude MLP} that predicts the Drude frequency, $\overline{\omega}_\mathrm{D}$.
The number of hidden layers in both output MLPs is again fixed to two.
The \textsc{Spectra MLP} produces a 4,002-dimensional output vector corresponding to the concatenated real and imaginary parts of $\overline{\varepsilon}_\mathrm{inter}(\omega)$, each sampled from $0$ to $20$~eV in $10$~meV steps, whereas the \textsc{Drude MLP} has a scalar output.
Based on past experience, we set the width of each hidden layer in the \textsc{Spectra MLP} to $m_\mathrm{Spectra}\times d_\mathrm{h}$, with $m_\mathrm{Spectra}=4$.
We tested several alternative values of $m_\mathrm{Spectra}$ for the \textsc{Spectra MLP} using the optimized TC-based \textsc{OptiMetal2B} architecture (see Supplementary Note 4) and found that performance remains unchanged apart from the expected scaling with model size (Supplementary Fig.~\ref{fig:spectra_dim}).

\newpage
\clearpage

\section*{Supplementary Note 4: Model architecture optimization}

To identify a well-performing architecture for the task at hand, we systematically evaluated the architectural choices introduced in Supplementary Note 3, using \textsc{Optuna} to tune optimizer hyperparameters for each candidate architecture \cite{Akiba2019}.
For each model type (\textsc{OptiMetal2B} or \textsc{OptiMetal3B}) and each network component (node embeddings, edge embeddings, message-passing layers, and pooling layers), we substituted one candidate variant at a time into a fixed base architecture, while keeping all remaining components unchanged.
A complete list of the investigated architecture choices and their corresponding hyperparameters is provided in Supplementary Tabs.~\ref{tab:node}, \ref{tab:edge}, \ref{tab:mp}, and \ref{tab:pool} for the node embeddings, edge embeddings, message-passing layers, and pooling layers, respectively.

\begin{table}[ht]
    \centering
    \caption{
    \textbf{Candidate node-embedding layers and associated hyperparameters considered during architecture optimization.}
    Boldface indicates the configuration selected as optimal.
    Implementation details for the listed layer types are provided in the Code Availability statement in the main text.
    }
    \vspace{2mm}
    \begin{tabular}{ll@{\hspace{1em}}l@{\hspace{1em}}}
    \toprule
    Layer type & Hyperparameter & Choices \\ \midrule
    Atomic number embedding & Embedding dimension $N_\mathrm{emb}$ & $[16, 32, 64, 128, 256, 512]$\\ \midrule
    \textbf{One-hot group $\|$ One-hot period} \cite{Grunert2024}
    ~~
    & --- & ---  \\ \bottomrule
    \end{tabular}
    \label{tab:node}
\end{table}

\begin{table}[ht]
    \centering
    \caption{
    \textbf{Candidate edge-embedding layers and associated hyperparameters considered during architecture optimization.}
    Boldface indicates the configuration selected as optimal.
    Implementation details for the listed layer types are provided in the Code Availability statement in the main text.
    }
    \vspace{2mm}
    \begin{tabular}{ll@{\hspace{1em}}l@{\hspace{1em}}}
    \toprule
    Layer type & Hyperparameter & Choices \\ \midrule
    \multirow{3}{*}{\textbf{Gaussian}}
    ~~
    & Number of basis functions $N_\mathrm{b}$ & $[16, 32, \textbf{64}, 128, 256, 512]$ \\
    & Basis width $w_\mathrm{b}$ \cite{Fu2025} & $[0.5, 1.0, \textbf{2.0}, 3.0, 4.0]$ \\
    & Smooth polynomial envelope \cite{Gasteiger2020} & $[\mathrm{True}, \textbf{False}]$ \\ \midrule
    \multirow{3}{*}{Bessel} & Number of basis functions $N_\mathrm{b}$ & $[16, 32, 64, 128, 256, 512]$     \\
    & Trainable Bessel wave numbers \cite{Gasteiger2020} & $[\mathrm{True}, \mathrm{False}]$ \\
    & Smooth polynomial envelope \cite{Gasteiger2020} & $[\mathrm{True}, \mathrm{False}]$ \\ \bottomrule
    \end{tabular}
    \label{tab:edge}
\end{table}

\begin{table}[ht]
    \centering
    \caption{
    \textbf{Candidate message-passing layers and associated hyperparameters considered during architecture optimization.}
    Boldface indicates the configuration selected as optimal.
    Implementation details for the listed layer types are provided in the Code Availability statement in the main text.
    }
    \vspace{2mm}
    \begin{tabular}{ll@{\hspace{1em}}l@{\hspace{1em}}}
    \toprule
    Layer type & Hyperparameter & Choices \\ \midrule
    CGC \cite{Xie2018} & One-hidden-layer MLP width $m_\mathrm{MLP}$ & $[2, 4, 6]\times d_\mathrm{h}$ \\ \midrule
    \multirow{2}{*}{\textbf{GATC} \cite{Brody2021}} 
    ~~
    & One-hidden-layer MLP width $m_\mathrm{MLP}$ & $[2, \textbf{4}, 6]\times d_\mathrm{h}$ \\
    & Number of attention heads $N_\mathrm{head}$ & $[1, 2, \textbf{4}, 8, 16, 32]$ \\ \midrule
    \multirow{2}{*}{TC \cite{Thekumparampil2018}} & One-hidden-layer MLP width $m_\mathrm{MLP}$ & $[2, 4, 6]\times d_\mathrm{h}$ \\
    & Number of attention heads $N_\mathrm{head}$ & $[1, 2, 4, 8, 16, 32]$ \\ \bottomrule
    \end{tabular}
    \label{tab:mp}
\end{table}

\begin{table}[ht]
    \centering
    \caption{
    \textbf{Candidate pooling layers and associated hyperparameters considered during architecture optimization.}
    Boldface indicates the configuration selected as optimal.
    Implementation details for the listed layer types are provided in the Code Availability statement in the main text.
    }
    \vspace{2mm}
    \begin{tabular}{ll@{\hspace{1em}}l@{\hspace{1em}}}
    \toprule
    Layer type & Hyperparameter & Choices \\ \midrule
    \multicolumn{1}{l}{Mean} & --- & --- \\ \midrule
    Scalar attention & --- & --- \\ \midrule
    \textbf{Vector attention} \cite{Grunert2024} 
    ~~
    & --- & --- \\ \midrule
    Set2Set \cite{Vinyals2015} & Iterations & $[1,2,3,4]$ \\ \bottomrule
    \end{tabular}
    \label{tab:pool}
\end{table}

During the architecture optimization, we set the hidden dimension to $d_\mathrm{h}=256$ and used a subset of the training set, consisting of 20,000 materials (see Supplementary Note 6), to reduce the computational costs.
All models were trained using the training protocol described in the Methods section of the main text.

Each candidate architecture was evaluated using a two-stage workflow that decouples architectural choices from optimizer hyperparameters.

\begin{figure}[ht]
    \centering
    \includegraphics{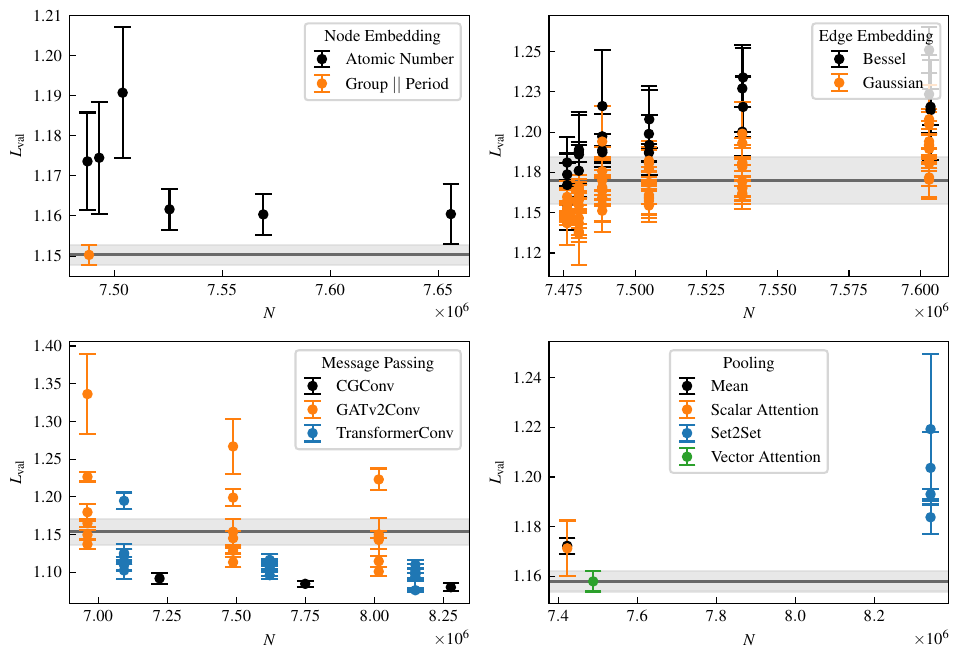}
    \caption{
    \textbf{Summary of the architecture optimization of \textsc{OptiMetal2B}.}
    The validation loss $L_\mathrm{val}$ is shown as a function of the number of trainable parameters, $N$, for all evaluated node embeddings, edge embeddings, message-passing layers, and pooling layers, as indicated by the panel legends.
    During the architecture optimization, one network component is varied, while the others remain fixed to the components used in \textsc{OptiMate} \cite{Grunert2024, Grunert2025}.
    Dots represent $L_\mathrm{val}$ averaged over three random model initializations, and error bars indicate the standard deviation.
    In each panel, the horizontal line and shaded band indicate the mean and standard deviation, respectively, of the reference \textsc{OptiMate}-based configuration.
    }
    \label{fig:arch_2b}
\end{figure}

In the first stage, we optimized the maximum learning rate, $\eta_\mathrm{max}$, and weight decay, $\lambda$, for a fixed reference random seed using \textsc{Optuna} \cite{Akiba2019} with a tree-structured Parzen estimator (TPE) sampler \cite{Watanabe2023} and median pruning.
We searched over a discrete grid of maximum learning rates, $\eta_\mathrm{max} \in \{10^{-5}, 2\times 10^{-5}, 4\times 10^{-5}, \dots, 8\times 10^{-3}\}$, and weight decays, $\lambda \in \{0, 10^{-6}, 2\times 10^{-6}, 4\times 10^{-6}, \dots, 10^{-3}\}$. 
A total of 25 trials were performed to identify the optimal optimizer hyperparameters $(\eta_\mathrm{max}, \lambda)$, each consisting of a short 100-epoch training run.
In the first ten trials, $(\eta_\mathrm{max}, \lambda)$ were sampled randomly and training was performed without pruning.
Subsequent trials used $(\eta_\mathrm{max}, \lambda)$ proposed by the TPE sampler, with pruning decisions evaluated every 25 epochs based on performance relative to the median trial.
The hyperparameter pair $(\eta_\mathrm{max}, \lambda)$, which resulted in the lowest validation loss, was chosen for the subsequent step.

In the second stage, each architecture was retrained from scratch three times for 200 epochs using different random initializations and the optimal optimizer hyperparameters $(\eta_\mathrm{max}, \lambda)$ obtained in the first stage.
The mean of the resulting minimum validation losses was used to compare architectural choices.

Using this workflow, we first optimized the architecture of \textsc{OptiMetal2B} by varying one network component at a time, while keeping the others fixed to the \textsc{OptiMate}-based reference \cite{Grunert2024, Grunert2025} to identify the best-performing layer variants.
In total, we evaluated 137 distinct \textsc{OptiMetal2B} architectures.
The results are summarized in Supplementary Fig.~\ref{fig:arch_2b}.
Performance improvements across all network components except the message-passing layer show little correlation with the total number of trainable parameters, $N$, indicating that architectural choices dominate model size in these cases.
In contrast, for the message-passing layers, we observe three distinct clusters along the $N$-axis, each of which corresponds to a fixed value of the one-hidden-layer MLP width multiplier, $m_\mathrm{MLP}$ (cf.~Supplementary Note 3).
Within each cluster, the message-passing layers are consistently ordered by performance from worst to best: GATC, TC, and then CGC.
For a given message-passing layer, increasing $m_\mathrm{MLP}$ results in a modest yet consistent improvement in validation performance, as evidenced by the corresponding cluster shifting toward lower validation loss.
This is further illustrated in Supplementary Fig.~\ref{fig:arch_2b_mp}, which shows the same data color-coded by $m_\mathrm{MLP}$.
While increasing $m_\mathrm{MLP}$ leads to modest performance improvements for a given message-passing layer, this trend primarily reflects raw parameter scaling rather than qualitative architectural changes, whereas the choice of the underlying message aggregation scheme has a substantial impact on the validation loss.

\begin{figure}[ht]
    \centering
    \includegraphics{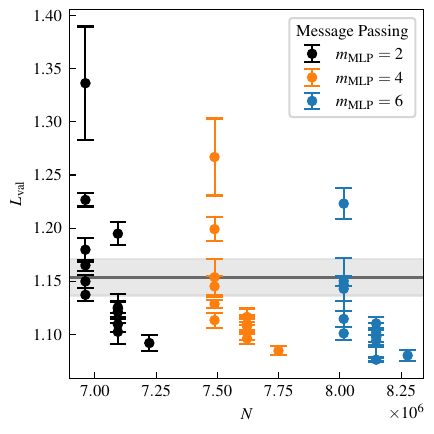}
    \caption{
    \textbf{Comparison of message-passing layers in \textsc{OptiMetal2B} with different coloration.}
    The validation loss $L_\mathrm{val}$ is shown as a function of the total number of trainable parameters $N$ for different message-passing layers.
    During the architecture optimization, one network component is varied, while the others remain fixed to the components used in \textsc{OptiMate} \cite{Grunert2024, Grunert2025}.
    The data points form three distinct clusters corresponding to fixed values of the one-hidden-layer MLP width multiplier $m_\mathrm{MLP}=2,4,6$.
    Within each cluster, the message-passing layers are consistently ordered by performance as \textsc{GATv2Conv} (GATC), \textsc{TransformerConv} (TC), and \textsc{CGConv} (CGC).
    Dots represent $L_\mathrm{val}$ averaged over three random model initializations, and error bars indicate the standard deviation.
    The horizontal line and shaded band indicate the mean and standard deviation, respectively, of the reference \textsc{OptiMate}-based configuration.
    }
    \label{fig:arch_2b_mp}
\end{figure}

Next, we assessed potential cross-layer interactions by combining the two best-performing choices of the node embedding, message-passing layer, and pooling layer, given in Supplementary Tab.~\ref{tab:top2}.
For the edge embedding, only the best-performing variant was retained, as the top two choices differed only in their use of a smooth polynomial envelope and exhibited nearly identical performance.
This yielded a total of eight architectural combinations, which we evaluated using the aforementioned two-stage optimization workflow.

\begin{table}[ht]
    \centering
    \caption{
    \textbf{Top-two architectural choices per network component identified in the layer-wise optimization of \textsc{OptiMetal2B}.}
    All permutations of these architectural components---excluding the second-best edge embedding---are combined to create candidate architectures for evaluating cross-layer interactions.
    }
    \vspace{2mm}
    \begin{tabular}{ll@{\hspace{1em}}l@{\hspace{1em}}}
    \toprule
    Layer type & Best & Second best \\ \midrule
    Node embedding & One-hot group $\|$ One-hot period & Atomic number embedding ($N_\mathrm{emb}=256$)  \\
    Edge embedding 
    ~~
    & Gaussian ($N_\mathrm{b}=32$, $w_\mathrm{b}=4.0$, $\mathrm{Envelope}=\mathrm{True}$) & Gaussian ($N_\mathrm{b}=32$, $w_\mathrm{b}=4.0$, $\mathrm{Envelope}=\mathrm{False}$) \\
    Message passing & TC ($N_\mathrm{head}=2$, $m_\mathrm{MLP}=6$) & CGC ($m_\mathrm{MLP}=6$) \\
    Pooling & Vector attention & Scalar attention \\
    \bottomrule
    \end{tabular}
    \label{tab:top2}
\end{table}

\begin{figure}[ht]
    \centering
    \includegraphics{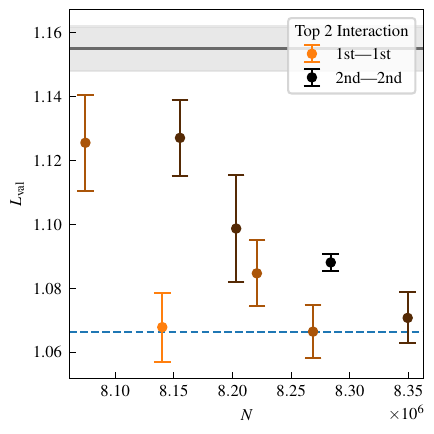}
    \caption{
    \textbf{Assessment of cross-layer interactions in \textsc{OptiMetal2B}.}
    The validation loss $L_\mathrm{val}$  is shown for each combination obtained by combining the two best-performing choices of each network component listed in Supplementary Tab.~\ref{tab:top2}.
    Dots represent $L_\mathrm{val}$ averaged over three random model initializations, and error bars indicate the standard deviation.
    The data points are colored continuously according to the number of components that are set to their best-performing variant.
    Black corresponds to a combination of all second-best-performing components, excluding the edge embedding. 
    Orange corresponds to a combination of all the best-performing components.
    The horizontal line and shaded band indicate the mean and standard deviation, respectively, of the reference \textsc{OptiMate}-based configuration.
    The dashed blue line indicates the configuration with the lowest validation loss, which performs slightly better than the combination of all the best-performing components.
    }
    \label{fig:arch_2b_cross_inter}
\end{figure}

As shown in Supplementary Fig.~\ref{fig:arch_2b_cross_inter}, the validation performance varies across the different combinations, showing no clear trend when transitioning from a configuration consisting entirely of second-best-performing components to one consisting entirely of the best-performing components.
To determine the origin of these variations, we examined the effect of individual architectural choices while keeping all other components fixed.
For example, starting from the best-performing configuration in Supplementary Tab.~\ref{tab:top2} and replacing vector-attention pooling with scalar-attention pooling increases the average validation loss from $1.068$ to $1.126$.
Similarly, replacing the group-period embedding with an atomic number embedding increases the loss from $1.068$ to $1.085$.
In contrast, replacing the top-performing TC message passing with the second-best-performing CGC message passing only reduces the validation loss by about $10^{-3}$.
These results indicate that the observed variations in performance are primarily due to the contributions of individual architectural components rather than strong cross-layer effects.
Although cross-layer interactions are present (cf.~TC versus CGC message passing), their magnitude appears weak within the explored search space, and the contributions of the individual architectural components are essentially additive.

For the 1D NSL analysis presented in the main text, we selected the \textsc{OptiMetal2B} variant with the best performance, as determined by the one-layer-at-a-time optimization (see Supplementary Tab.~\ref{tab:top2}), with all architectural components fixed to their respective optimal choices.
As the cross-layer interaction study revealed, replacing the TC message-passing layer with the second-best-performing CGC variant (with $m_\mathrm{MLP}=6$), while keeping all other components unchanged, yields a lower validation loss, though only by about $10^{-3}$.
Therefore, we included this CGC-based architecture in the 1D NSLs to examine the effect of the message-passing layer on the NSLs explicitly.

For \textsc{OptiMetal3B}, we adopted the final \textsc{OptiMetal2B} configuration as the starting point for the architecture optimization.
Angles were embedded using spherical harmonics $Y_l^0$ up to $l=3$, resulting in a four-dimensional angle embedding.
We fixed the node and edge embeddings, as well as the pooling layer, to the configurations that performed best in \textsc{OptiMetal2B} (see Supplementary Tab.~\ref{tab:top2}), and restricted the architecture optimization to the \textsc{Message Passing Blocks} inside the three-body \textsc{3B-Block}.
This restriction is motivated by two considerations.
First, it substantially reduces computational cost, as training \textsc{OptiMetal3B} is significantly more expensive than training \textsc{OptiMetal2B}.
Second, the cross-layer interaction results for \textsc{OptiMetal2B} tentatively suggest that optimal embeddings and pooling layers are largely transferable between the two architectures.
Since the primary structural extension in \textsc{OptiMetal3B} is the introduction of the \textsc{3B-Block}, it is therefore the natural focus of re-optimization.
Recall that the primary graph and the line graph use the same \textsc{Message Passing Blocks} in the \textsc{3B-Block}.

\begin{figure}[ht]
    \centering
    \includegraphics{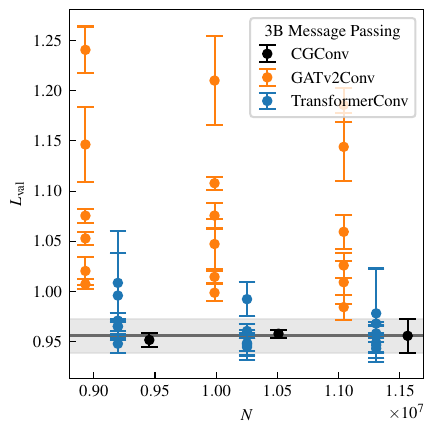}
    \caption{
    \textbf{Summary of the message-passing layer optimization of \textsc{OptiMetal3B}.}
    The validation loss $L_\mathrm{val}$ is shown as a function of the total number of trainable parameters $N$ for different message-passing layers.
    Dots represent $L_\mathrm{val}$ averaged over three random model initializations, and error bars indicate the standard deviation.
    In each panel, the horizontal line and shaded band indicate the mean and standard deviation, respectively, of the reference configuration based on the optimized \textsc{OptiMetal2B} architecture.
    Similar to Supplementary Fig.~\ref{fig:arch_2b_mp}, the results exhibit discrete clusters along the $N$-axis.
    Again, these clusters primarily arise from increases in the width multiplier of the one-hidden-layer MLP $m_{\mathrm{MLP}}$, and thus mainly reflect raw parameter scaling rather than qualitative architectural differences.
    }
    \label{fig:arch_3b_mp}
\end{figure}

Using this approach, we evaluated 39 different \textsc{Message Passing Blocks} in the \textsc{3B-Block}, again following the two-stage protocol described above.
The results are summarized in Supplementary Fig.~\ref{fig:arch_3b_mp}.
The optimal \textsc{Message Passing Block} for the final \textsc{OptiMetal3B} model employs TC-based message passing with eight attention heads and a one-hidden-layer MLP width of $m_\mathrm{MLP} = 6$.
In contrast to \textsc{OptiMetal2B}, the performance difference between TC- and CGC-based message passing is more pronounced in \textsc{OptiMetal3B}.
Compared to the optimized \textsc{OptiMetal2B} architecture, the corresponding optimized TC-based message-passing layer in \textsc{OptiMetal3B} employs a larger number of attention heads.
We tentatively interpret this observation as TC message passing benefiting from the richer geometric information introduced by explicit three-body interactions, which may enable different attention heads to specialize on distinct angular or bonding environments.

Combining the best-performing layers for \textsc{OptiMetal2B} yielded an overall performance improvement of about 8\% relative to the baseline \textsc{OptiMate}-style model.
For \textsc{OptiMetal3B} again, TC-based message passing yielded the best performance, though it only improved upon the base configuration with optimized embeddings and a pooling layer by 2\%.
Overall, the optimized \textsc{OptiMetal3B} achieves about 12\% better performance on the validation set than the optimized \textsc{OptiMetal2B}, demonstrating that incorporating three-body interactions significantly improves model performance.

\newpage
\clearpage

\section*{Supplementary Note 5: Model performance}

\begin{figure*}[ht]
    \centering
    \includegraphics{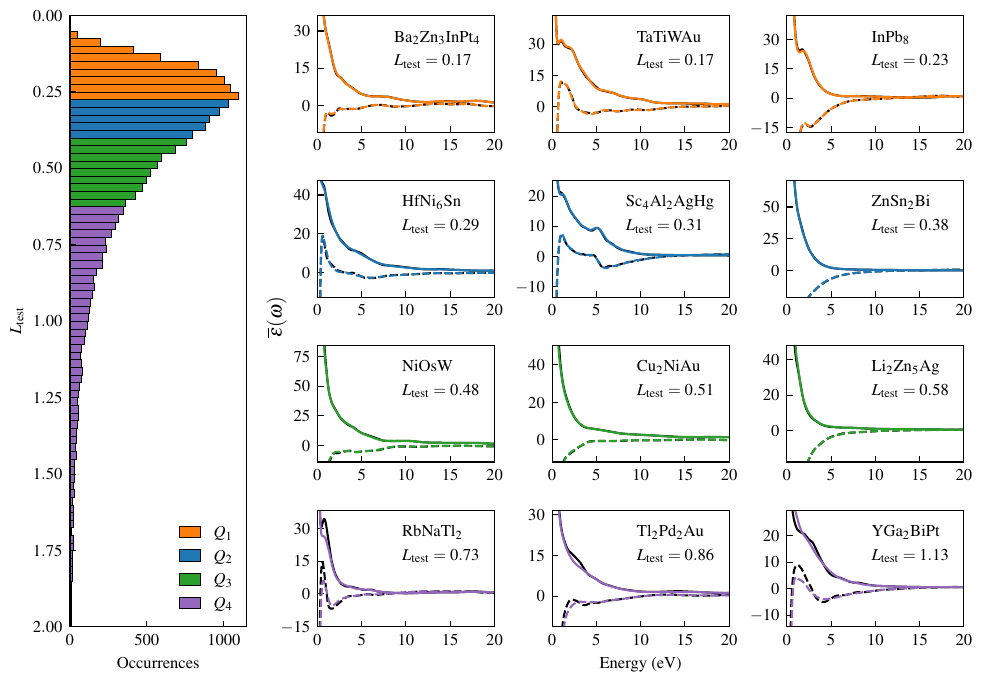}
    \caption{
    \textbf{Comparison of ab initio and ML-predicted total dielectric functions on the test set using an ensemble average of three \textsc{OptiMetal2B} models trained with different random initializations.}
    The large panel on the left shows the distribution of the test loss, $L_\mathrm{test}$, with quantiles highlighted in different colors every $25\%$.
    The panel grid on the right displays three randomly selected materials from each quantile.
    The ab initio total dielectric function is shown in black, while the ML-predicted ones are colored according to their quantile.
    The real part is shown as dashed lines, and the imaginary part as solid lines.
    The chemical composition and test loss of each material are indicated in the corresponding panel.
    }
    \label{fig:quant_2b}
\end{figure*}

\begin{figure*}[ht]
    \centering
    \includegraphics{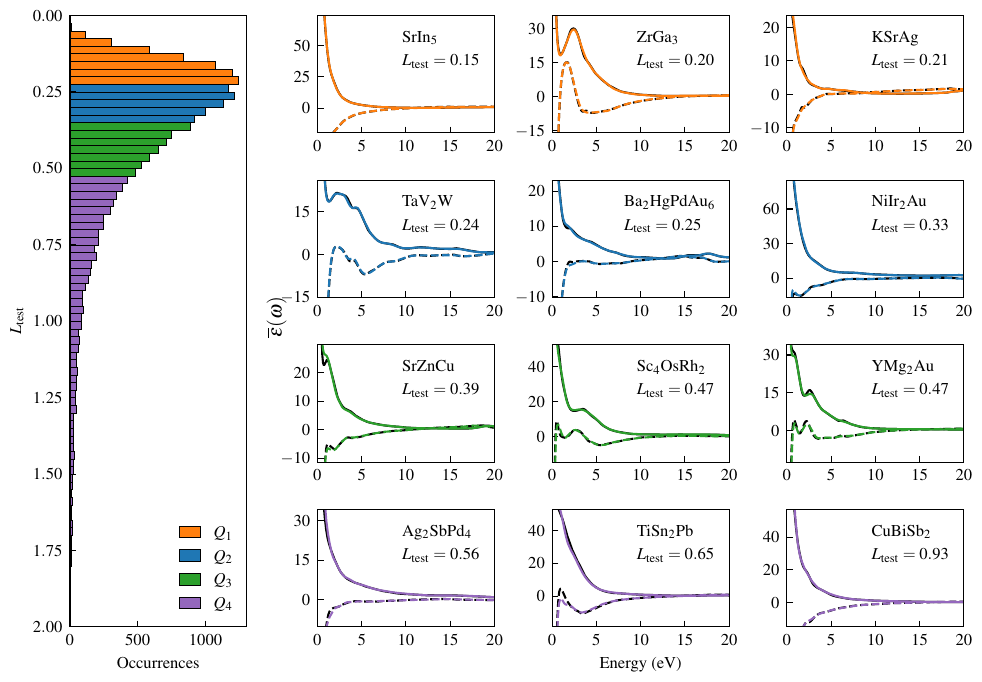}
    \caption{
    \textbf{Comparison of ab initio and ML-predicted total dielectric functions on the test set using an ensemble average of three \textsc{OptiMetal3B} models trained with different random initializations.}
    The large panel on the left shows the distribution of the test loss, $L_\mathrm{test}$, with quantiles highlighted in different colors every $25\%$.
    The panel grid on the right displays three randomly selected materials from each quantile.
    The ab initio total dielectric function is shown in black, while the ML-predicted ones are colored according to their quantile.
    The real part is shown as dashed lines, and the imaginary part as solid lines.
    The chemical composition and test loss of each material are indicated in the corresponding panel.
    }
    \label{fig:quant_3b}
\end{figure*}

In this Supplementary Note, we evaluate the performance of the \textsc{OptiMetal2B} and \textsc{OptiMetal3B} models with the optimized architectures identified in Supplementary Note 4 on the held-out test set.

For this, we selected the models used in the 2D NSL maps with a training set size of $D=160{,}000$ and width $d_\mathrm{h}=256$ ($N\approx 10\,\mathrm{M}$).
To obtain robust predictions, we trained three independent instances of each architecture with different random initializations, and then averaged the predicted observables across the ensemble.

Supplementary Fig.~\ref{fig:quant_2b} and \ref{fig:quant_3b} show the ensemble-averaged predictions of the selected \textsc{OptiMetal2B} and \textsc{OptiMetal3B} models, respectively, across the full range of test-set performance.
In each figure, the test-loss distribution, shown in the large panel on the left, is divided into four quantiles (where lower is better): i.e., 0--25\% (orange), 25--50\% (blue), 50--75\% (green), and 75--100\% (purple).
For each quantile, three representative materials are shown in the grid on the right, where the predicted total dielectric functions are colored accordingly.
Across all four quantiles in both figures, including the highest-loss quantile, the ML-predicted dielectric functions closely follow their ab initio counterparts, reproducing both global spectral trends and fine structural features over the full energy range.
This illustrates the ability of both \textsc{OptiMetal2B} and \textsc{OptiMetal3B} to accurately capture the frequency-dependent optical properties of metals.

\begin{table}[ht]
    \centering
    \caption{
    \textbf{Test-set performance metrics for \textsc{OptiMetal2B} and \textsc{OptiMetal3B}, evaluated using ensemble averages of three models trained with different random initializations.}
    The following metrics are reported: the test loss $L_\mathrm{test}$; the mean absolute error (MAE), coefficient of determination (R$^2$), and similarity coefficient (SC) \cite{Grunert2024} (see Methods in the main text) for the real and imaginary parts of the interband dielectric function $\overline{\varepsilon}_\mathrm{inter}(\omega)$; the absolute error (AE) and the absolute percentage error (APE) for the Drude frequency $\overline{\omega}_\mathrm{D}$; and the CIELAB color-difference metric $\Delta E$ (see text).
    For each metric, the table lists the mean, median, and standard deviation across all materials in the test set.
    }
    \vspace{2mm}
    \begin{tabular}{l@{\hspace{1em}}ccc@{\hspace{1.5em}}ccc}
    \toprule
     & \multicolumn{3}{@{\hspace{1em}}c@{\hspace{1.5em}}}{\textsc{OptiMetal2B}} & \multicolumn{3}{c}{\textsc{OptiMetal3B}} \\[1mm]
    Metric & Mean & Median & $\sigma$ & Mean & Median & $\sigma$ \\
    \midrule
    $L_\mathrm{test}$ & 0.496 & 0.385 & 0.392 & 0.425 & 0.335 & 0.317  \\
    \midrule
    MAE$[\mathrm{Re}(\overline{\varepsilon}_\mathrm{inter})]$ & 0.347 & 0.277 & 0.255 & 0.303 & 0.246 & 0.211 \\
    R$^2[\mathrm{Re}(\overline{\varepsilon}_\mathrm{inter})]$  & 0.979 & 0.993 & 0.056 & 0.984 & 0.995 & 0.036 \\
    SC$[\mathrm{Re}(\overline{\varepsilon}_\mathrm{inter})]$ & 0.892 & 0.913 & 0.073 & 0.905 & 0.923 & 0.064 \\
    \midrule
    MAE$[\mathrm{Im}(\overline{\varepsilon}_\mathrm{inter})]$  & 0.348 & 0.280 & 0.249 & 0.304 & 0.249 & 0.208 \\
    R$^2[\mathrm{Im}(\overline{\varepsilon}_\mathrm{inter})]$ & 0.970 & 0.990 & 0.092 & 0.978 & 0.992 & 0.058 \\
    SC$[\mathrm{Im}(\overline{\varepsilon}_\mathrm{inter})]$ & 0.930 & 0.944 & 0.049 & 0.939 & 0.950 & 0.042 \\
    \midrule
    AE$[\overline{\omega}_\mathrm{D}]$ (eV) & 0.149 & 0.088 & 0.189 & 0.121 & 0.074 & 0.150 \\
    APE$[\overline{\omega}_\mathrm{D}]$ (\%) & 3.523 & 1.957 & 5.791 & 2.898 & 1.636 & 4.720 \\
    \midrule
    $\Delta E$ & 1.521 & 1.109 & 1.584 & 1.365 & 0.992 & 1.402\\
    \bottomrule
    \end{tabular}
    \label{tab:metrics}
\end{table}

To quantitatively assess the accuracy of the selected \textsc{OptiMetal2B} and \textsc{OptiMetal3B} models, we report the mean, median, and standard deviation of several performance metrics across all materials in the test set in Supplementary Tab.~\ref{tab:metrics}.
Specifically, we evaluate the mean absolute error (MAE), the coefficient of determination (R$^2$), and the similarity coefficient (SC) \cite{Grunert2024} (see Methods in the main text), for the real and imaginary parts of the interband dielectric functions $\overline{\varepsilon}_\mathrm{inter}(\omega)$, as well as the absolute error (AE) and the absolute percentage error (APE) for the Drude frequency $\overline{\omega}_\mathrm{D}$.
In addition, we report the color difference metric $\Delta E$, defined---following the standards of the \textit{Commission Internationale de l'Eclairage} (CIE)---as the Euclidean distance in the approximately perceptually uniform CIELAB color space (CIE76).
Colors are computed from the total dielectric function $\overline{\varepsilon}(\omega)$ using the CIE standard illuminant D65 \cite{D65} and the CIE 1931 $2^\circ$ standard observer \cite{CIE1931}, following the work of Prandini et al.~\cite{Prandini2019Photo}.

Across the reported metrics, \textsc{OptiMetal3B} consistently outperforms \textsc{OptiMetal2B}, exhibiting a lower test loss and reduced MAE for both the real and imaginary parts of the interband dielectric function.
The differences in mean R$^2$ and mean SC are less pronounced, which is expected since both metrics are close to their optimal values, indicating that both models accurately capture the variance and overall shape of the ab initio reference spectra. 
Notably, SC values above $0.9$ reflect excellent spectral agreement, as this metric converges only slowly toward unity.

The mean MAE values for the interband response are larger than those reported for dielectric functions of semiconductors and insulators predicted using GNNs in previous studies \cite{Ibrahim2024, Grunert2024, Hung2024, Grunert2025, Grunert2026}.
This is somewhat expected, as the dielectric functions of intermetallic compounds tend to have higher-amplitude peaks and sharper spectral features, which are more severely penalized by absolute-error metrics.
Thus, the observed MAE values reflect the increased spectral complexity of metallic systems rather than poorer predictive accuracy compared to GNNs for spectra of semiconductors and insulators.

For the intraband contribution, the mean AE in the Drude frequency remains at or below $0.15$~eV for both models, corresponding to relative errors of only a few percent. 
This level of accuracy is consistent with the uncertainty of the ab initio calculations underlying the dataset (see Methods in the main text).

The mean CIELAB color difference $\Delta E$ remains below $1.6$ for both models, which is smaller than the commonly reported just-noticeable-difference threshold in CIELAB ($\Delta E \approx 2.3$) \cite{Sharma1997}, indicating that the residual discrepancies between ML-predicted and ab initio spectra are, in most cases, imperceptible to the human eye.

These results demonstrate that the improved scaling behavior of \textsc{OptiMetal3B} translates into systematic gains in predictive accuracy.
Nevertheless, \textsc{OptiMetal2B} already achieves a high level of accuracy when trained on the large dataset produced in this study.

\newpage
\clearpage

\section*{Supplementary Note 6: Composition of subsampled training datasets}

\begin{figure}[ht]
    \centering
    \includegraphics{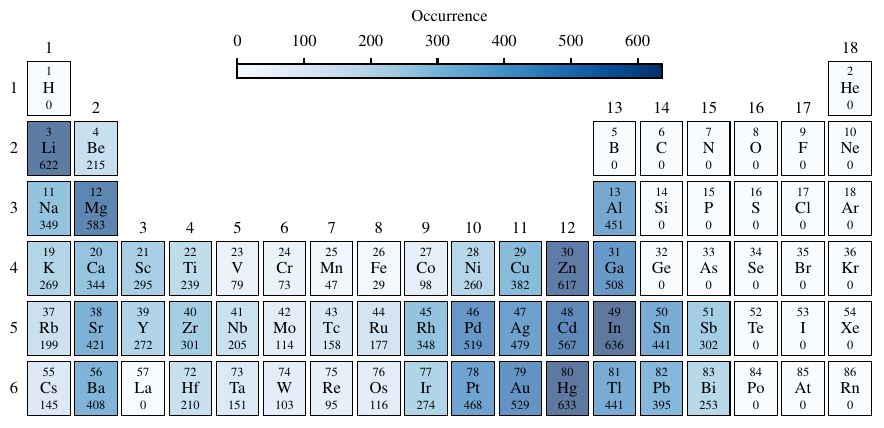}
    \caption{
    \textbf{Periodic table illustrating the elemental distribution in the subsampled training set of 2,500 materials.}
    Colors indicate the number of occurrences of each element in the dataset, with exact counts shown below the respective symbols.
    The lanthanides and the seventh period are omitted, as there are no elements from these groups in the dataset.
    }
    \label{fig:pse2500}
\end{figure}

\begin{figure}[ht]
    \centering
    \includegraphics{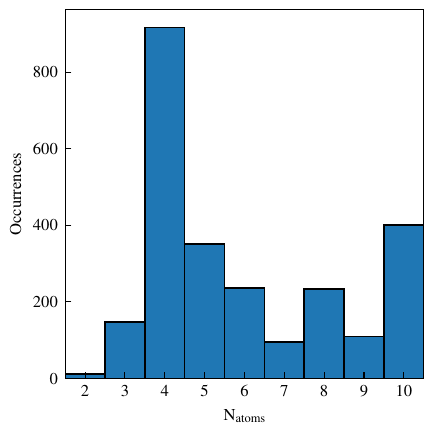}
    \caption{
    \textbf{Distribution of the number of atoms per unit cell for all compounds in the subsampled training set of 2,500 materials.}
    }
    \label{fig:nsites2500}
\end{figure}

\begin{figure}[ht]
    \centering
    \includegraphics{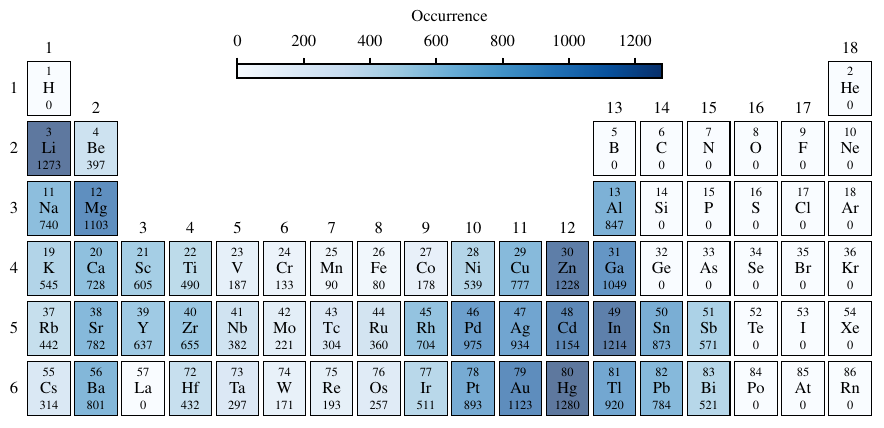}
    \caption{
    \textbf{Periodic table illustrating the elemental distribution in the subsampled training set of 5,000 materials.}
    For more details, see Supplementary Fig.~\ref{fig:pse2500}.
    }
    \label{fig:pse5000}
\end{figure}

\begin{figure}[ht]
    \centering
    \includegraphics{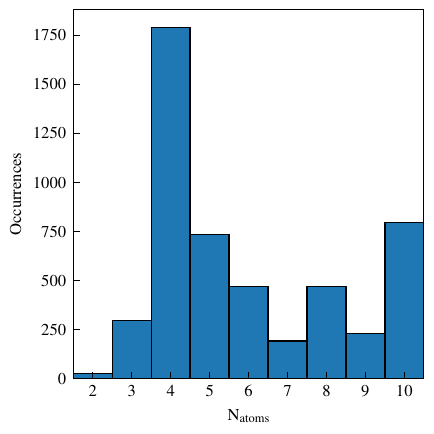}
    \caption{
    \textbf{Distribution of the number of atoms per unit cell for all compounds in the subsampled training set of 5,000 materials.}
    }
    \label{fig:nsites5000}
\end{figure}

\begin{figure}[ht]
    \centering
    \includegraphics{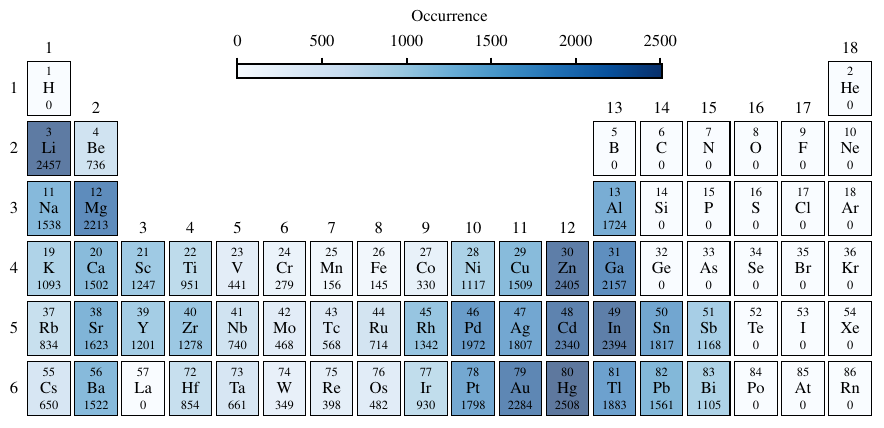}
    \caption{
    \textbf{Periodic table illustrating the elemental distribution in the subsampled training set of 10,000 materials.}
    For more details, see Supplementary Fig.~\ref{fig:pse2500}.
    }
    \label{fig:pse10000}
\end{figure}

\begin{figure}[ht]
    \centering
    \includegraphics{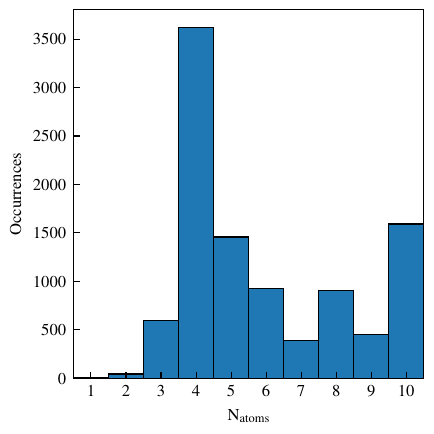}
    \caption{
    \textbf{Distribution of the number of atoms per unit cell for all compounds in the subsampled training set of 10,000 materials.}
    }
    \label{fig:nsites10000}
\end{figure}

\begin{figure}[ht]
    \centering
    \includegraphics{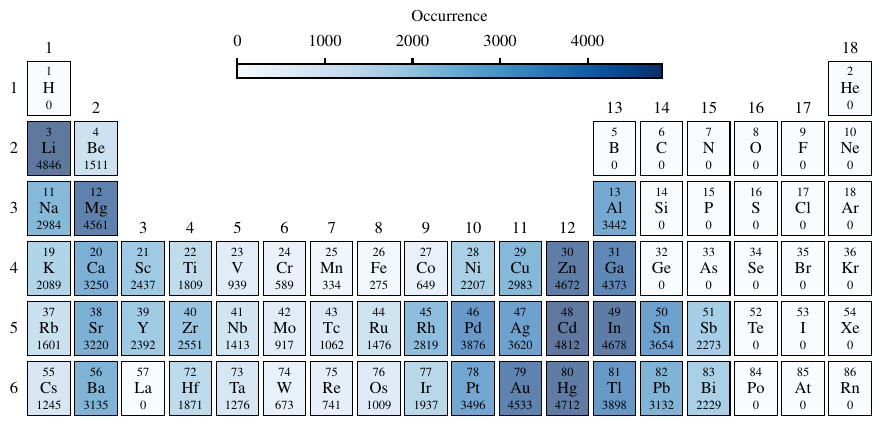}
    \caption{
    \textbf{Periodic table illustrating the elemental distribution in the subsampled training set of 20,000 materials.}
    For more details, see Supplementary Fig.~\ref{fig:pse2500}.
    }
    \label{fig:pse20000}
\end{figure}

\begin{figure}[ht]
    \centering
    \includegraphics{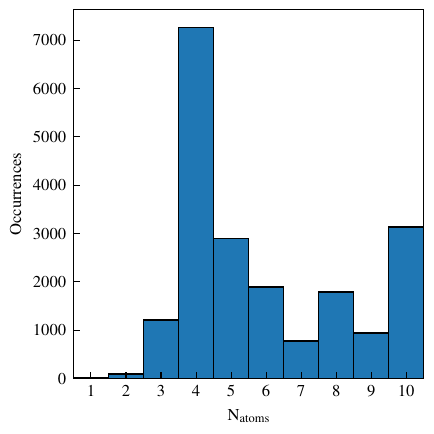}
    \caption{
    \textbf{Distribution of the number of atoms per unit cell for all compounds in the subsampled training set of 20,000 materials.}
    }
    \label{fig:nsites20000}
\end{figure}

\begin{figure}[ht]
    \centering
    \includegraphics{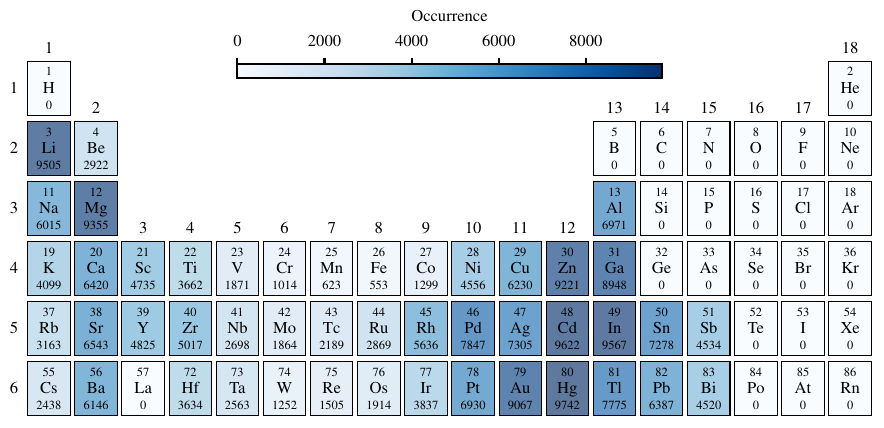}
    \caption{
    \textbf{Periodic table illustrating the elemental distribution in the subsampled training set of 40,000 materials.}
    For more details, see Supplementary Fig.~\ref{fig:pse2500}.
    }
    \label{fig:pse40000}
\end{figure}

\begin{figure}[ht]
    \centering
    \includegraphics{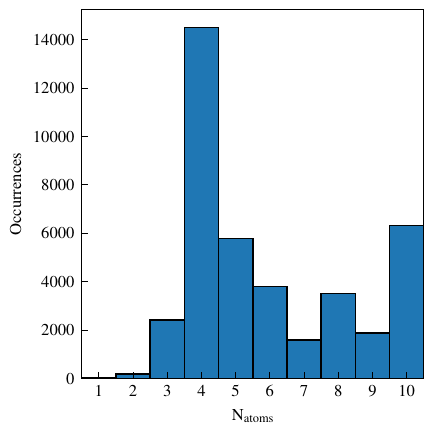}
    \caption{
    \textbf{Distribution of the number of atoms per unit cell for all compounds in the subsampled training set of 40,000 materials.}
    }
    \label{fig:nsites40000}
\end{figure}

\begin{figure}[ht]
    \centering
    \includegraphics{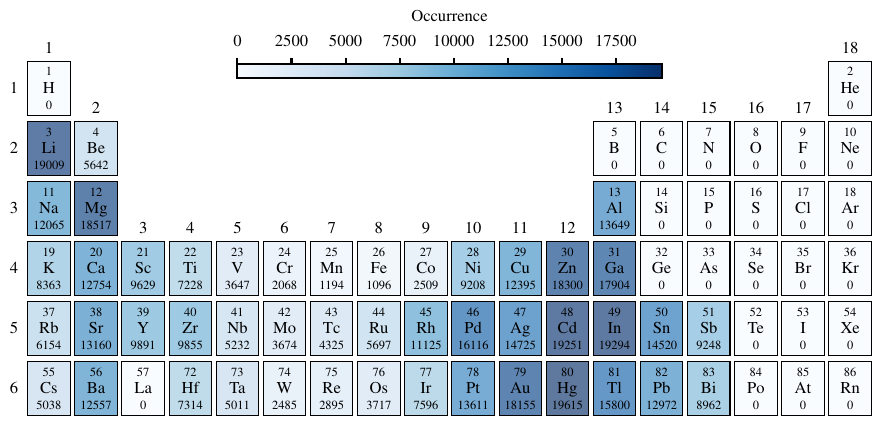}
    \caption{
    \textbf{Periodic table illustrating the elemental distribution in the subsampled training set of 80,000 materials.}
    For more details, see Supplementary Fig.~\ref{fig:pse2500}.
    }
    \label{fig:pse80000}
\end{figure}

\begin{figure}[ht]
    \centering
    \includegraphics{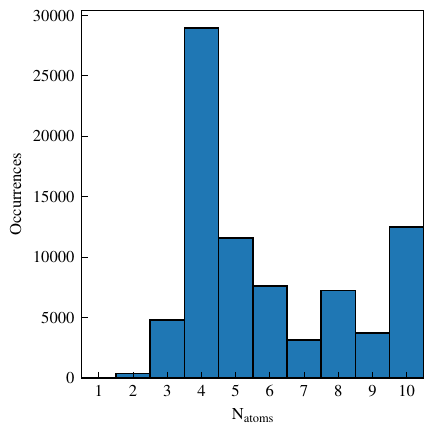}
    \caption{
    \textbf{Distribution of the number of atoms per unit cell for all compounds in the subsampled training set of 80,000 materials.}
    }
    \label{fig:nsites80000}
\end{figure}

\begin{figure}[ht]
    \centering
    \includegraphics{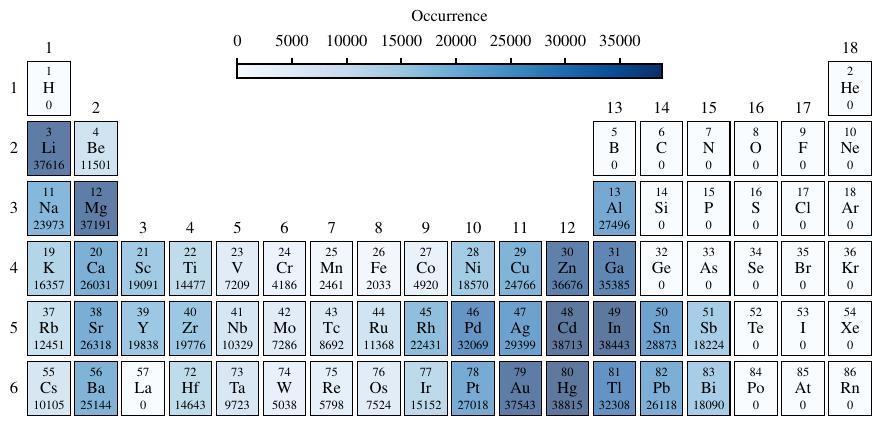}
    \caption{    
    \textbf{Periodic table illustrating the elemental distribution in the subsampled training set of 160,000 materials.}
    For more details, see Supplementary Fig.~\ref{fig:pse2500}.
    }
    \label{fig:pse160000}
\end{figure}

\begin{figure}[ht]
    \centering
    \includegraphics{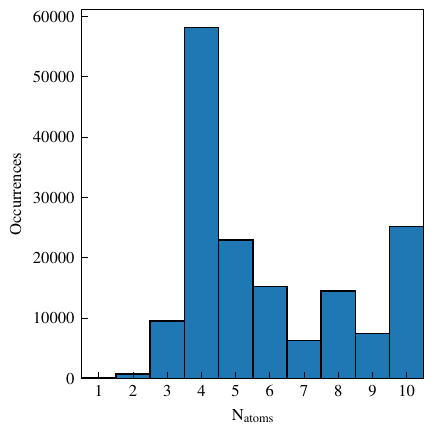}
    \caption{
    \textbf{Distribution of the number of atoms per unit cell for all compounds in the subsampled training set of 160,000 materials.}
    }
    \label{fig:nsites160000}
\end{figure}

\newpage
\clearpage

\section*{Supplementary Note 7: AICc tables for NSL fits}

This Supplementary Note reports the Akaike information criterion corrected for small sample sizes (AICc, see main text and Ref.~\cite{Burnham2002}) values for all functional forms considered in the 1D and 2D NSL analyses presented in the main text.
The AICc provides an objective measure for comparing competing fits, balancing goodness of fit with model complexity. 
Lower values indicate a preferred description of the data.

For scaling with dataset size, $D$, the obtained AICc values for all considered NSL functional forms are given in Supplementary Tab.~\ref{tab:1d_data_aicc}.
Among these, the smoothly broken power law without an additional amplitude parameter consistently yields the lowest AICc and is therefore selected as the preferred functional form.

Examining the AICc values for scaling with respect to the number of model parameters, $N$, in Supplementary Tab.~\ref{tab:1d_parameter_aicc}, the power law with a saturation floor yields the lowest AICc for \textsc{OptiMetal2B} (CGC) and \textsc{OptiMetal3B} (TC).
However, for \textsc{OptiMetal2B} (TC), the smoothly broken power law with an additional amplitude parameter achieves a slightly lower AICc, though this improvement is marginal and likely arises from the increased flexibility introduced by the additional amplitude parameter.
To enable consistent comparisons of scaling behavior across all models and message-passing formulations, we therefore adopt the power law with a saturation floor for \textsc{OptiMetal2B} (TC) as well.

Supplementary Tab.~\ref{tab:2d} reports the AICc values for the 2D NSL fits, $L(D,N)$, comparing the smoothly interpolating formulation inspired by Kaplan et al.~\cite{Kaplan2020} with the additive formulation following Hoffmann et al.~\cite{Hoffmann2022}.
For both \textsc{OptiMetal2B} (TC) and \textsc{OptiMetal3B} (TC), the Kaplan-style NSL achieves substantially lower AICc values, indicating a significantly better trade-off between goodness of fit and model complexity.
These results indicate that $L(D,N)$ is better described by a coupled, smoothly interpolating scaling law than by an additive formulation.
Accordingly, the Kaplan-style NSL is adopted as the preferred 2D NSL functional form in the main text.

\begin{table}[ht]
    \centering
    \footnotesize
    \caption{
    \textbf{AICc values for 1D NSL fits as a function of dataset size.}
    For each model and message-passing formulation, four functional forms were fitted to the validation loss $L_\mathrm{val}(D)$: a power law, a power law with a saturation floor, a smoothly broken power law, and a smoothly broken power law with an amplitude parameter.
    Lower AICc values indicate a better trade-off between goodness of fit and model complexity.
    Boldface highlights the preferred functional form for each model.
    The referenced equations can be found in the Methods section of the main text.
    }
    \vspace{2mm}
    \begin{tabular}{lc@{\hspace{1em}}c@{\hspace{1em}}c@{\hspace{1em}}}
    \toprule
     & \textsc{OptiMetal2B} (CGC) & \textsc{OptiMetal2B} (TC) & \textsc{OptiMetal3B} (TC) \\
    \midrule
    Power law, cf.~Eq.~(8) & $-36.20$ & $-34.87$ & $-39.91$ \\
    Power law with saturation floor, cf.~Eq.~(9) & $-29.20$ & $-27.87$ & $-32.91$ \\
    Smoothly broken power law, cf.~Eq.~(10) & {$\mathbf{-57.74}$}& $\mathbf{-44.88}$ & $\mathbf{-57.70}$ \\
    Smoothly broken power law with amplitude, cf.~Eq.~(11) & $-49.89$ & $-38.72$ & $-48.48$ \\
    \bottomrule
    \end{tabular}
    \label{tab:1d_data_aicc}
\end{table}

\begin{table}[ht]
    \centering
    \footnotesize
    \caption{
    \textbf{AICc values for 1D NSL fits as a function of the number of model parameters.}
    For each model and message-passing formulation, four functional forms were fitted to the validation loss $L_\mathrm{val}(N)$: a power law, a power law with a saturation floor, a smoothly broken power law, and a smoothly broken power law with an amplitude parameter (see Methods in the main text).
    Lower AICc values indicate a better trade-off between goodness of fit and model complexity.
    Boldface highlights the preferred functional form for each model.
    The referenced equations can be found in the Methods section of the main text.
    }
    \vspace{2mm}
    \begin{tabular}{lc@{\hspace{1em}}c@{\hspace{1em}}c@{\hspace{1em}}}
    \toprule
     & \textsc{OptiMetal2B} (CGC) & \textsc{OptiMetal2B} (TC) & \textsc{OptiMetal3B} (TC) \\
    \midrule
    Power law, cf.~Eq.~(8) & $-41.67$ & $-27.99$ & $-44.40$ \\
    Power law with saturation floor, cf.~Eq.~(9) & $\mathbf{-60.97}$ & $\mathbf{-54.73}$ & $\mathbf{-51.17}$ \\
    Smoothly broken power law, cf.~Eq.~(10) & $-48.73$ & $\mathbf{-56.10}$ & $-47.24$ \\
    Smoothly broken power law with amplitude, cf.~Eq.~(11) & $-50.34$ & $-44.71$ & $-47.04$ \\
    \bottomrule
    \end{tabular}
    \label{tab:1d_parameter_aicc}
\end{table}

\begin{table}[ht]
    \centering
    \caption{
    \textbf{AICc values for 2D NSL fits.}
    For each model, two functional forms were fitted to the validation loss $L_\mathrm{val}(D, N)$: a smoothly interpolating model inspired by Kaplan et al.~\cite{Kaplan2020} and an additive formulation following Hoffmann et al.~\cite{Hoffmann2022}.
    Lower AICc values indicate a better trade-off between goodness of fit and model complexity.
    Boldface highlights the preferred functional form for each model.
    The referenced equations can be found in the Methods section of the main text.
    }
    \vspace{2mm}
    \begin{tabular}{lc@{\hspace{1em}}c@{\hspace{1em}}}
    \toprule
     & \textsc{OptiMetal2B} (TC) & \textsc{OptiMetal3B} (TC) \\
    \midrule
    Kaplan-style NSL, cf.~Eq.~(14) & $\mathbf{-344.50}$ & $\mathbf{-359.55}$ \\
    Hoffmann-style NSL, cf.~Eq.~(15) & $-253.68$ & $-285.21$  \\
    \bottomrule
    \end{tabular}
    \label{tab:2d}
\end{table}

\newpage
\clearpage

\section*{Supplementary Note 8: Convergence of optical properties with respect to the Brillouin zone sampling}

\begin{figure*}[ht]
    \centering
    \includegraphics{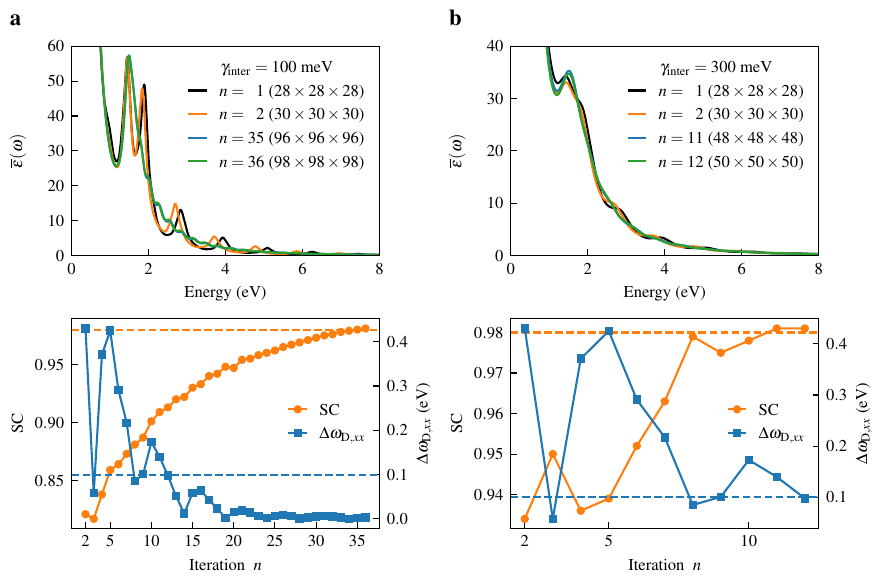}
    \caption{
    \textbf{Illustration of the convergence behavior of the total dielectric function and Drude frequency of aluminum with respect to the Brillouin zone sampling.}
    \textbf{a} Convergence behavior with $\gamma_\mathrm{inter}=100$~meV.
    \textbf{b} Convergence behavior with $\gamma_\mathrm{inter}=300$~meV.
    The upper panels show the rotationally invariant total dielectric function,
    $\overline{\varepsilon}(\omega)$, computed on the first and last two k-point grids encountered in the third step of the high-throughput workflow described in the Methods section of the main text.
    The lower panels show the corresponding convergence behavior of the similarity coefficient (SC), defined in Eq.~(5), and of the $xx$-component of the Drude frequency tensor, $\omega_{\mathrm{D}, xx}$, as the k-point grid is iteratively refined.
    Note the markedly different SC scales of both panels.
    Convergence is reached once the SC between consecutive grids exceeds $0.98$ (evaluated between $1$ and $20$~eV) and the change in $\omega_{\mathrm{D},xx}$ falls below $100$~meV.
    The convergence thresholds are marked by dashed horizontal lines in the lower panels.
    }
    \label{fig:al_convergence}
\end{figure*}

Supplementary Fig.~\ref{fig:al_convergence} illustrates the convergence of the total dielectric function and Drude frequency of aluminum with respect to the Brillouin zone sampling for two different values of the interband broadening, $\gamma_\mathrm{inter}$, following the workflow described in the Methods section of the main text.
The crystal structure used for aluminum is the same as that employed in the experimental validation described in Supplementary Note 2 (cf.~Supplementary Tab.~\ref{tab:simple_metals}).
This example demonstrates that the choice of $\gamma_\mathrm{inter}$ has a significant impact on the convergence behavior.
With a smaller broadening of $\gamma_\mathrm{inter}=100$~meV, convergence is achieved only with an extremely dense $98 \times 98 \times 98$ k-point grid.
In contrast, for $\gamma_\mathrm{inter}=300$~meV, convergence is reached with a comparatively coarser  $50 \times 50 \times 50$ k-point grid.
Moreover, as shown in Supplementary Fig.~\ref{fig:exp1} in Supplementary Note 2, an interband broadening of $\gamma_\mathrm{inter}=300$~meV yields good agreement with experimental amplitudes and peak shapes.
Therefore, based on these computational considerations and the benchmarking results discussed in Supplementary Note 2, we adopted $\gamma_\mathrm{inter}=300$~meV throughout the high-throughput workflow.
We note, however, that some materials, such as elemental beryllium, still require extremely dense k-point grids to converge the Drude frequency within the $100$~meV convergence threshold.

\newpage
\clearpage

\section*{Supplementary Note 9: Composition of dataset splits}

\begin{figure}[ht]
    \centering
    \includegraphics{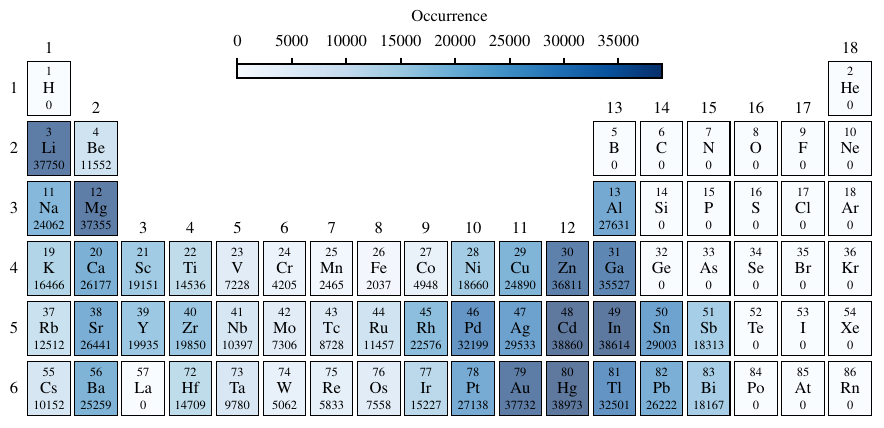}
    \caption{
    \textbf{Periodic table illustrating the elemental distribution in the training set.}
    Colors indicate the number of occurrences of each element in the dataset, with exact counts shown below the respective symbols.
    The lanthanides and the seventh period are omitted, as there are no elements from these groups in the dataset.
    }
    \label{fig:pse_train}
\end{figure}

\begin{figure}[ht]
    \centering
    \includegraphics{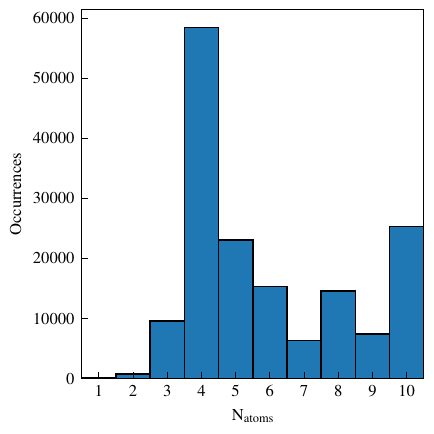}
    \caption{
    \textbf{Distribution of the number of atoms per unit cell for all compounds in the training set.}
    }
    \label{fig:nsites_train}
\end{figure}

\begin{figure}[ht]
    \centering
    \includegraphics{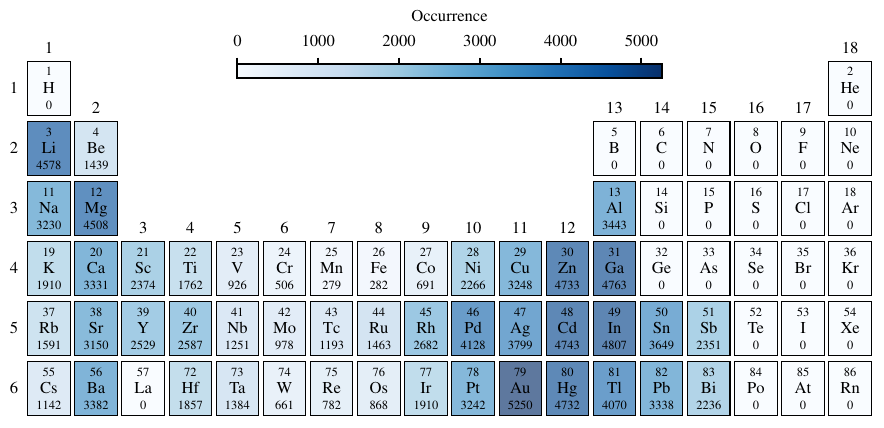}
    \caption{
    \textbf{Periodic table illustrating the elemental distribution in the validation set.}
    Colors indicate the number of occurrences of each element in the dataset, with exact counts shown below the respective symbols.
    The lanthanides and the seventh period are omitted, as there are no elements from these groups in the dataset.
    }
    \label{fig:pse_val}
\end{figure}

\begin{figure}[ht]
    \centering
    \includegraphics{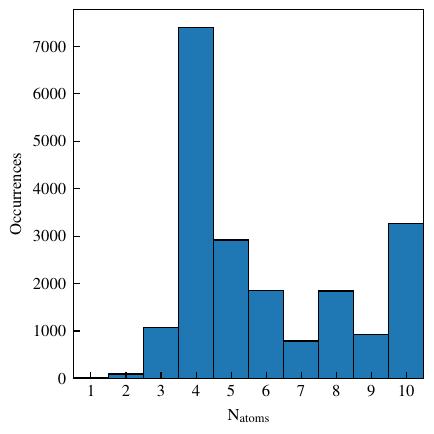}
    \caption{
    \textbf{Distribution of the number of atoms per unit cell for all compounds in the validation set.}
    }
    \label{fig:nsites_val}
\end{figure}

\begin{figure}[ht]
    \centering
    \includegraphics{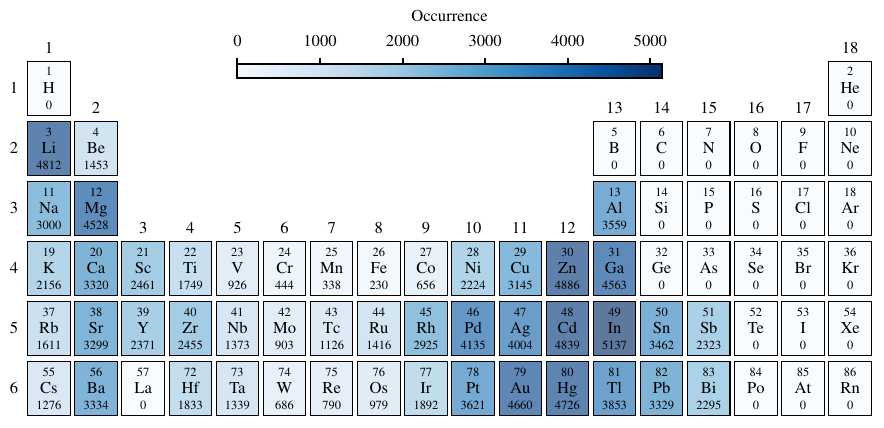}
    \caption{
    \textbf{Periodic table illustrating the elemental distribution in the test set.}
    Colors indicate the number of occurrences of each element in the dataset, with exact counts shown below the respective symbols.
    The lanthanides and the seventh period are omitted, as there are no elements from these groups in the dataset.
    }
    \label{fig:pse_test}
\end{figure}

\begin{figure}[ht]
    \centering
    \includegraphics{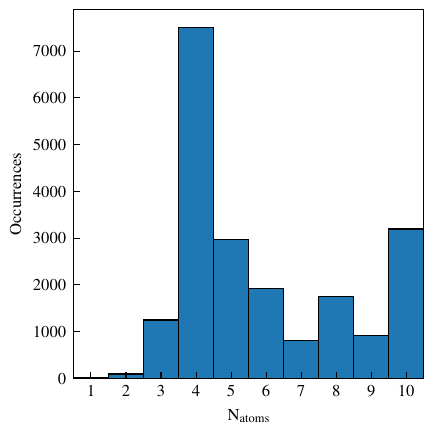}
    \caption{
    \textbf{Distribution of the number of atoms per unit cell for all compounds in the test set.}
    }
    \label{fig:nsites_test}
\end{figure}

\newpage
\clearpage

\section*{Supplementary Note 10: Learning rate scaling for TC message passing}

\begin{figure}[ht]
    \centering
    \includegraphics{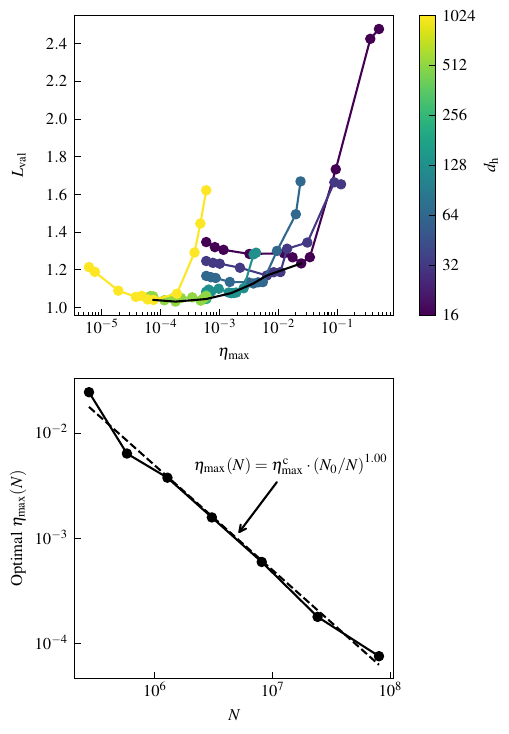}
    \caption{
    \textbf{Learning-rate scaling for TC-based message passing.}
    The upper panel shows the validation loss $L_\mathrm{val}$ as a function of the maximum learning rate $\eta_\mathrm{max}$ for TC-based \textsc{OptiMetal2B} at different hidden dimensions $d_\mathrm{h}$ (color bar).
    All models were trained for $500$ epochs using the subset of the training set consisting of 20,000 materials (see Supplementary Note 6), with all architectural components and optimizer hyperparameters, except $d_\mathrm{h}$ and $\eta_\mathrm{max}$, fixed to the values selected in Supplementary Note 4.
    Dots represent $L_\mathrm{val}$ averaged over three random model initializations.
    For each width, the optimal learning rate is defined as the $\eta_\mathrm{max}$ that minimizes the validation loss averaged over three random model initializations across the sweep, indicated by the black line connecting the per-width minima.
    The lower panel shows the empirically optimal learning rates $\eta_\mathrm{max}(N)$ as a function of the number of trainable parameters $N$.
    The dashed line corresponds to a power-law fit of the form $\eta_\mathrm{max}(N)=\eta_{\mathrm{max}}^{c}(N_0/N)^{\gamma}$, where $N_0$ denotes the parameter count of the reference model with $d_\mathrm{h}=256$.
    }
    \label{fig:tc_scaling}
\end{figure}

As mentioned in the Methods section of the main text, we found that training of TC-based models became unstable at hidden dimensions $d_\mathrm{h} > 256$ when using the maximum learning rate, $\eta_\mathrm{max}$, identified at $d_\mathrm{h} = 256$ in Supplementary Note 4.
Therefore, during the NSL analyses, we deliberately avoided re-optimizing the maximum learning rate for each model width, ensuring that the observed NSLs reflected dependence on the number of trainable parameters alone.
Instead, we adopted a learning-rate scaling strategy in which $\eta_\mathrm{max}$ is adjusted as a function of the number of model parameters, $N$.
To determine an appropriate learning-rate scaling, we varied $\eta_\mathrm{max}$ for different $d_\mathrm{h}$ (i.e., $N$) for the optimized TC-based \textsc{OptiMetal2B} model (see Supplementary Note 4).
The results are shown in Supplementary Fig.~\ref{fig:tc_scaling}.
In the upper panel, the optimal learning rate for $d_\mathrm{h}$ is identified as the value of $\eta_\mathrm{max}$ that minimizes the validation loss averaged over three random model initializations.
As shown in the lower panel, the optimal $\eta_\mathrm{max}$ depends on the total number of trainable parameters, $N$. 
This dependency is well described by a power-law scaling of the form $\eta_\mathrm{max}(N)\propto N^{-\gamma}$.
Based on these results, we set $\gamma=1$, with $N_0$ set to the number of trainable parameters of the model with $d_\mathrm{h}=256$ for which $\eta_\mathrm{max}$ was optimized in Supplementary Note 4.
This choice ensures stable training at large $N$ in the NSL experiments in the main text, eliminating the need to re-optimize $\eta_\mathrm{max}$ for each model width and allowing us to isolate pure parameter scaling.

\newpage
\clearpage

\bibliography{literature_si}